\documentclass[usegraphicx,usenatbib]{mn2e}

\usepackage{amsmath}
\usepackage{amssymb}
\usepackage{color}
\title[Colliding clusters and DM self-interactions]{Colliding clusters and dark matter self-interactions}
\author[F.~Kahlhoefer, K.~Schmidt-Hoberg, M.~T.~Frandsen and S.~Sarkar]{Felix Kahlhoefer,$^1$\thanks{felix.kahlhoefer@physics.ox.ac.uk} Kai Schmidt-Hoberg,$^2$ Mads T.~Frandsen$^3$ and Subir Sarkar$^{1,4}$ \\
$^1$ Rudolf Peierls Centre for Theoretical Physics, University of Oxford, 1 Keble Road, Oxford OX1 3NP, United Kingdom\\
$^2$ Theory Division, CERN, 1211 Geneva 23, Switzerland\\
$^3$ CP$^3$-Origins and the Danish Institute for Advanced Study, University of Southern Denmark, Campusvej 55, 5230 Odense M, Denmark\\
$^4$ Niels Bohr Institute, Blegdamsvej 17, 2100 K{\o}benhavn {\O}, Denmark}

\begin{document}

\maketitle

\begin{abstract}
When a dark matter halo moves through a background of dark matter particles, self-interactions can lead to both deceleration and evaporation of the halo and thus shift its centroid relative to the collisionless stars and galaxies. We study the magnitude and time evolution of this shift for two classes of dark matter self-interactions, viz.~frequent self-interactions with small momentum transfer (e.g.~due to long-range interactions) and rare self-interactions with large momentum transfer (e.g.~contact interactions), and find important differences between the two cases. We find that neither effect can be strong enough to completely separate the dark matter halo from the galaxies, if we impose conservative bounds on the self-interaction cross-section. The majority of both populations remain bound to the same gravitational potential and the peaks of their distributions are therefore always coincident. Consequently any \emph{apparent} separation is mainly due to particles which are leaving the gravitational potential, so will be largest shortly after the collision but not observable in evolved systems. Nevertheless the fraction of collisions with large momentum transfer is an important characteristic of self-interactions, which can potentially be extracted from observational data and provide an important clue as to the nature of dark matter.
\end{abstract}

\section{Introduction}

The successful standard paradigm for structure formation on cosmological scales assumes that dark matter (DM) is collisionless and cold. Indeed in most models of particle DM such as supersymmetric theories, the DM self-interaction cross-section is comparable to the DM-nucleon scattering cross-section, which is experimentally constrained to be extremely small. However there is no reason why the dark sector cannot be strongly coupled to itself, as long as the interactions with Standard Model (SM) particles are sufficiently weak. This setup would be natural in models with a rich dark sector which has new gauge forces~\citep{Carlson:1992fn, Mohapatra:2001sx, Kusenko:2001vu, PhysRevD.84.051703}.

The possibility of large self-interactions in the dark sector was suggested as a solution to the well-known problems of the standard cold dark matter (CDM) cosmology on galactic scales~\citep{1995ApJ...452..495D,Spergel:1999mh}. Subsequently, various bounds on the self-interaction cross-section have been derived from the study of different astrophysical systems~\citep{Yoshida:2000uw,0004-637X-564-1-60,Firmani:2000qe,Gnedin:2000ea,Dave:2000ar,Hennawi:2001be,Markevitch:2003at,2005ApJ...631..244S,Boehm:2005,Randall:2007ph}. These bounds constrain the simplest models for DM self-interactions (e.g.~contact interactions with velocity-independent cross-section) to the point where they may no longer be sufficient to reduce the tension at small scales (see~\citet{Rocha:2012jg,Peter:2012jh,Vogelsberger:2012ku,Zavala:2012us} for a recent discussion). Invoking a (plausible) velocity dependence, however, one can evade these bounds~\citep{Ackerman:2008gi,Feng:2009mn,Buckley:2009in,Loeb:2010gj,Aarssen:2012fx,Tulin:2013teo}.

Any evidence for DM self-interactions would have striking implications for particle physics, as it would severely constrain or even rule out popular candidates such as supersymmetric neutralinos and axions. A sensitive probe for this purpose is a DM halo moving through a larger system with a high background density of DM particles. This could e.g.~be a satellite of the Milky Way moving through the Galactic halo, or a galaxy with its DM halo moving inside a galaxy cluster. Systems of particular interest in this context are colliding galaxy clusters, such as the `Bullet Cluster'~\citep{Markevitch:2003at,Clowe:2006eq}, Abell 520~\citep{Mahdavi:2007yp, Jee:2012sr} or the recently discovered `Musket Ball Cluster'~\citep{2012ApJ...747L..42D, Dawson:2012fx}. Independently of the difficulty in understanding how such systems can arise in the standard CDM cosmology with initial gaussian perturbations~\citep{Farrar:2006tb, Lee:2010hja, Watson:2013mea}, one can immediately derive a bound on the DM self-interaction cross-section from the simple observation that such systems have survived the collision and not evaporated as a result of the energy transferred from the larger system.

A potentially more sensitive probe of DM self-interactions is whether these have caused the DM halo to slow down. In particular, such a deceleration can lead to an observable separation between the DM halo and the collisionless stars or galaxies~\citep{Markevitch:2003at,Randall:2007ph}. The aim of this paper is to develop an intuitive understanding of the origin of this separation from a particle physics point of view based on simple analytical arguments and to confirm our expectations with numerical simulations of colliding clusters. We find that even in the presence of DM self-interactions, the peak of the DM distribution always remains \emph{coincident} with the peak of the distribution of stars/galaxies. However, self-interactions can induce an \emph{asymmetry} in the two distributions due to particles (either DM particles or stars/galaxies) which escape from the combined gravitational potential or travel on highly elliptical orbits. This asymmetry results in a separation of the respective centroids. However, most DM particles and galaxies will remain bound to the same gravitational potential, which has important implications for the magnitude and time-dependence of the separation. Our findings agree with the observation of small but typically non-zero offsets between the DM halo and the brightest cluster galaxy in 10,000 Sloan Digital Sky Survey clusters~\citep{2012MNRAS.426.2944Z}.

Our central observation is that the momentum transfer cross-section $\sigma_\mathrm{T}$ of DM self-interactions is insufficient to completely characterise the behaviour of the system and the properties of the separation. This is because the same $\sigma_\text{T}$ can arise both from frequent collisions with small momentum transfer, or from rare interactions with large momentum transfer. In systems with a strong directionality, such as two colliding clusters, the angular distribution of the scattered DM particles plays a particularly important role. As an additional parameter to characterise DM self-interactions, we introduce the fraction of expulsive collisions $f$, which quantifies the probability for collisions with large momentum transfer. Only if $f$ is much smaller than unity, it is possible to have frequent DM self-interactions without violating observational constraints on the evaporation rate. If this is the case, DM self-interactions can be described by an effective drag force. On the other hand, if $f$ is large, DM self-interactions must be rare and an effective description of collective effects is impossible, because only some fraction of the DM particles will be affected at all. This important distinction has often been neglected in the interpretation of observational data. In particular, many numerical simulations of self-interacting DM assume rare self-interactions, while many analytical arguments are based on the assumption of an effective drag force.

Our paper is structured as follows: In \S~\ref{sec:setup}, we introduce our notation and present the general setup of the problem. In particular, we define the fraction of expulsive collisions and discuss its relation to the evaporation rate. We provide a precise definition of the separation between DM haloes and stars or galaxies and discuss whether such a separation can arise from gravitational interactions alone. In \S~\ref{sec:freqint} we turn to frequent self-interactions and calculate the resulting drag force on the DM halo. We then compare these analytical estimates with the results of our numerical simulations. We provide additional information and detailed calculations related to this section in Appendices~\ref{ap:evaporation} and \ref{ap:frequent}. The case of contact interactions is discussed in \S~\ref{sec:rare}. Even though an effective description of such interactions is not possible, we develop qualitative arguments as well as a simple analytical model to predict the separation. Again, using an extended numerical simulation, we confirm our expectations. Further details are provided in Appendices~\ref{ap:numerical} and \ref{ap:analytical}.

\section{General setup}
\label{sec:setup}

We consider a gravitationally bound system $S_1$ moving in the gravitational potential of a larger system $S_2$. For example, $S_1$ can be a Milky Way satellite with $S_2$ being the Milky Way or $S_1$ is a galaxy moving in a galaxy cluster $S_2$. We will be most interested in the case of collisions of clusters and will use $S_1$ to denote the smaller cluster (called sub-cluster) and $S_2$ to denote the larger cluster (called main cluster). For this reason, we will always assume that $S_1$ is composed of (self-interacting) DM and collisionless galaxies. The case where $S_1$ is composed of DM and stars (instead of galaxies) can be treated in complete analogy. The crucial property of all these cases is that the typical velocities of particles in $S_1$ (i.e.~the velocity dispersion $\sigma_1$ and escape velocity $v_{\text{esc},1}$) are \emph{much smaller} than the relative velocity between the two systems $\bmath{v}_0$. We always choose our coordinate system to be centred at $S_2$ such that $\bmath{v}_0$ points in $z$-direction. Consequently, $z$ measures the distance between $S_1$ and $S_2$. In the following we focus on the evolution of $S_1$ and $S_2$ after their closest approach, so $z$ increases with time.

\begin{figure*}
\begin{center}
\includegraphics[width=0.53 \textwidth]{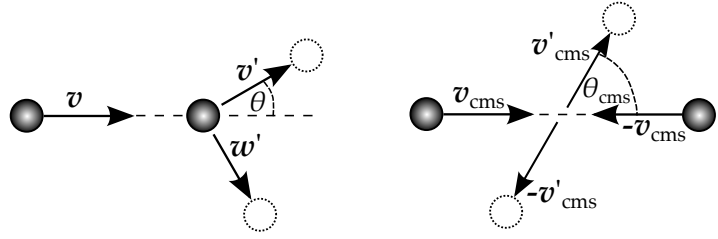}
\vspace{1mm}
\caption{\label{fig:variables}Collision between two DM particles in the rest frame of one of the particles (left) and in the centre-of-mass frame (right).}
\end{center}
\end{figure*}

We denote the differential scattering cross-section for DM self-interactions by $\mathrm{d}\sigma/\mathrm{d}\Omega_\text{cms}$ in the centre-of-mass frame and by $\mathrm{d}\sigma/\mathrm{d}\Omega$ in
the laboratory frame (which will be the rest frame of one of the two particles). Before the collision, the velocities of the two DM particles are denoted by $\bmath{v}$ and $\bmath{w}$ ($\bmath{v}_\text{cms}$ and $\bmath{w}_\text{cms}$) in the laboratory frame (centre-of-mass frame). The corresponding quantities after the collision are $\bmath{v}'$ and $\bmath{w}'$ ($\bmath{v}'_\text{cms}$ and $\bmath{w}'_\text{cms}$). Note that $\bmath{w}_\text{cms} = - \bmath{v}_\text{cms}$ and $\bmath{w}'_\text{cms} = - \bmath{v}'_\text{cms}$. We use $\theta$ and $\theta_\text{cms}$ to denote the scattering angle between incoming and outgoing DM particle (note that $\theta_\text{cms} = 2 \theta$). We assume that the two DM particles are indistinguishable, so the differential cross-section must be symmetric under the exchange $\theta_\text{cms} \rightarrow \pi - \theta_\text{cms}$.\footnote{To avoid complications, we assume here that there is no distinction between DM particles and antiparticles, or that DM is asymmetric i.e.~antiparticles are absent altogether.}
These kinematic variables are illustrated in Fig.~\ref{fig:variables}.

In the literature, most attention has been paid to the case where $d\sigma/\mathrm{d}\Omega_\text{cms}$ is \emph{independent} of the scattering angle, i.e.~isotropic scattering. Such an isotropic scattering results e.g.~from the exchange of a heavy particle which can be integrated out to give an effective point-like interaction.\footnote{Note that the DM particles are non-relativistic ($v \ll 1$), so effective point-like interactions will arise as soon as the mediator mass satisfies $m_\text{med} \gg v \, m_\text{DM}$.} However when the mediator of the interaction is lighter than the momentum exchange in the scattering, the differential cross-section will almost always have a strong dependence on $\theta_\text{cms}$. For a massless mediator, the differential cross-section typically diverges in the limit $\theta_\text{cms} \rightarrow 0$ and $\theta_\text{cms} \rightarrow \pi$. Such long-range interactions arise naturally in models of Mirror DM~\citep{Blinnikov:1983gh, Kolb:1985bf, Berezhiani:1995am, Foot:2004pa} or Atomic DM~\citep{Kaplan:2009de, Cline:2012is, CyrRacine:2012fz}.

\subsection{Halo evaporation}

Large DM self-interactions can lead to the evaporation of DM haloes~\citep{Gnedin:2000ea,Markevitch:2003at}. Here, we distinguish two different mechanisms that can lead to evaporation. \emph{Immediate} evaporation (denoted by subscript ``imd'') occurs if in a single collision the momentum transfer is large enough to remove a DM particle from the halo. We refer to such collisions as \emph{expulsive}. On the other hand, \emph{cumulative} evaporation (denoted by subscript ``cml'') can result from a large number of non-expulsive collisions, if in each of these collisions a DM particle gains a small amount of energy.

Let us begin by calculating the rate of expulsive collisions.\footnote{We return to the rate of cumulative evaporation in \S~\ref{sec:freqint} and Appendix~\ref{ap:evaporation}.} For this purpose, it will be most convenient to work in the frame where $S_1$ is at rest and $S_2$ moves at a velocity $\bmath{v}_0$. An immediate evaporation occurs if both $v'$ and $w' = \sqrt{v^2 + w^2 - v'^2}$ are larger than $v_\text{esc,1}$ at the position where the collision occurs: $v'^2 > v_\text{esc,1}^2$ and $v^2 + w^2 - v'^2 > v_\text{esc,1}^2$.

For the moment, we make the simplifying assumption that the relative velocity between the two systems is large compared to their individual velocity dispersions, so that we can approximate $\bmath{v} \approx 0$ and $\bmath{w} \approx \bmath{v}_0$. We then find $v' = v_0 \cos \theta = v_0 \sqrt{(1 + \cos \theta_\text{cms}) / 2}$. Moreover, we neglect the position dependence of the escape velocity $v_\text{esc,1}$. The rate of immediate evaporation, defined as $R_\text{imd} = N^{-1} \,\mathrm{d}N_\text{imd}/\mathrm{d}t$ (where $N$ is the total number of
DM particles in $S_1$) is then given approximately by
\begin{equation}
R_\text{imd} = \frac{\rho_2}{m_\text{DM}} \, v_0 \int \mathrm{d}\phi_\text{cms} \int_{2 \, v_\text{esc,1}^{2}/v_0^2 - 1}^{1-2 \, v_\text{esc,1}^{2}/v_0^2} \mathrm{d} \cos \theta_\text{cms} \, \frac{\mathrm{d}\sigma}{\mathrm{d}\Omega_\text{cms}} \; ,
\end{equation}
where $\rho_2 / m_\text{DM}$ is the number density of DM particles in
$S_2$ and $\mathrm{d}\Omega_\text{cms} = \mathrm{d}\phi_\text{cms} \, \mathrm{d} \cos \theta_\text{cms}$. Integrating $R_\text{imd}$ along the path of $S_1$ we obtain the
total fraction of evaporated DM particles
\begin{equation}
\frac{\Delta N_\text{imd}}{N} = 1 - \exp\left( - \frac{\Sigma_2}{m_\text{DM}} \,
\int_{2 \, v_\text{esc,1}^{2}/v_0^2 - 1}^{1-2 \, v_\text{esc,1}^{2}/v_0^2} \mathrm{d}\Omega_\text{cms}
\, \frac{\mathrm{d}\sigma}{\mathrm{d}\Omega_\text{cms}} \right)\; ,
\end{equation}
where $\Sigma_2 = \int \rho_2(z)\,\mathrm{d}z$ is the integrated density.

The expected number of DM particles lost because of immediate evaporation can be used to constrain DM self-interactions (see e.g.~\citet{Markevitch:2003at}). Here, we will be particularly interested in the fraction of expulsive collisions, given by:
\begin{equation}
f = \frac{\int_{2 \, v_\text{esc,1}^{2}/v_0^2 - 1}^{1 - 2 \, v_\text{esc,1}^{2}/v_0^2} \mathrm{d}\Omega_\text{cms} \, \left(\mathrm{d}\sigma/\mathrm{d}\Omega_\text{cms}\right)}{\int \mathrm{d}\Omega_\text{cms} \, \left(\mathrm{d}\sigma/\mathrm{d}\Omega_\text{cms}\right)} \; .
\end{equation}
For the case of contact interactions, this reduces to~\citep{Markevitch:2003at}:
\begin{equation}
f = 1 - \frac{2 \, v_\text{esc}^2}{v_0^2} \equiv 1 - \kappa \; .
\end{equation}
With this definition of the fraction of expulsive collisions, $f$, we can write
\begin{equation}
\frac{\Delta N_\text{imd}}{N} = 1 - \exp \left[ - \frac{\Sigma_2 \, \sigma \, f}{m_\text{DM}} \right] \; ,
\end{equation}
where $\sigma \equiv \int \mathrm{d}\Omega_\text{cms} \, \mathrm{d}\sigma/\mathrm{d}\Omega_\text{cms}$ is the total scattering cross-section. We observe that the fraction of evaporated DM particles is large if either $f \approx 1$ and $\Sigma_2 \, \sigma / m_\text{DM} \approx 1$, or if $f \ll 1$ but $\Sigma_2 \, \sigma / m_\text{DM} \gg 1$. The first case corresponds to the case of \emph{rare} self-interactions with high probability of expulsive collisions, the second corresponds to the case of \emph{frequent} self-interactions with low probability of expulsive collisions.

As far as the evaporation rate is concerned, these two kinds of interactions are largely indistinguishable. However, as we will see below, when characterising the separation between DM haloes and galaxies arising from self-interactions we obtain fundamentally \emph{different} predictions for rare self-interactions with $f \approx 1$ and frequent self-interactions with $f \ll 1$. We will therefore make a clear distinction between these two cases from now on.

Before beginning our discussion of separation, let us consider a simple example. For the Bullet Cluster $v_\text{esc,1} \approx 1900\,\text{km\,s}^{-1}$ and $v_0 \approx 4500\, \text{km\,s}^{-1}$, so $\kappa \equiv 2 \, v_\text{esc}^2 / v_0^2 \approx 0.4$. Consequently, expulsive collisions occur whenever $50\degr \la \theta_\text{cms} \la 130\degr$\footnote{For a larger value of $\theta$ the incoming particles loses so much momentum that it becomes bound to $S_1$, so the roles of the two particles will be exchanged.} which for isotropic scattering gives a fraction of expulsive collisions of $f \approx 0.6$. This large value implies that in order for \emph{isotropic} DM scattering to not immediately destroy the sub-cluster, it must be rare, i.e.~the majority of DM particles must not have scattered even once as the sub-cluster passed through the main cluster. If, on the other hand, we want to consider frequent DM self-interactions, meaning the average DM particle scatters many times during the cluster collision, we must require that expulsive collisions be rare, i.e.~$f \ll 1$. In other words, we must then require that the differential cross-section be strongly peaked at $\cos \theta_\text{cms} = \pm 1$ in order to have a sufficiently small rate for immediate evaporation.

\subsection{Separation between haloes and galaxies}
\label{sec:separation}

\begin{figure}
\begin{center}
\includegraphics{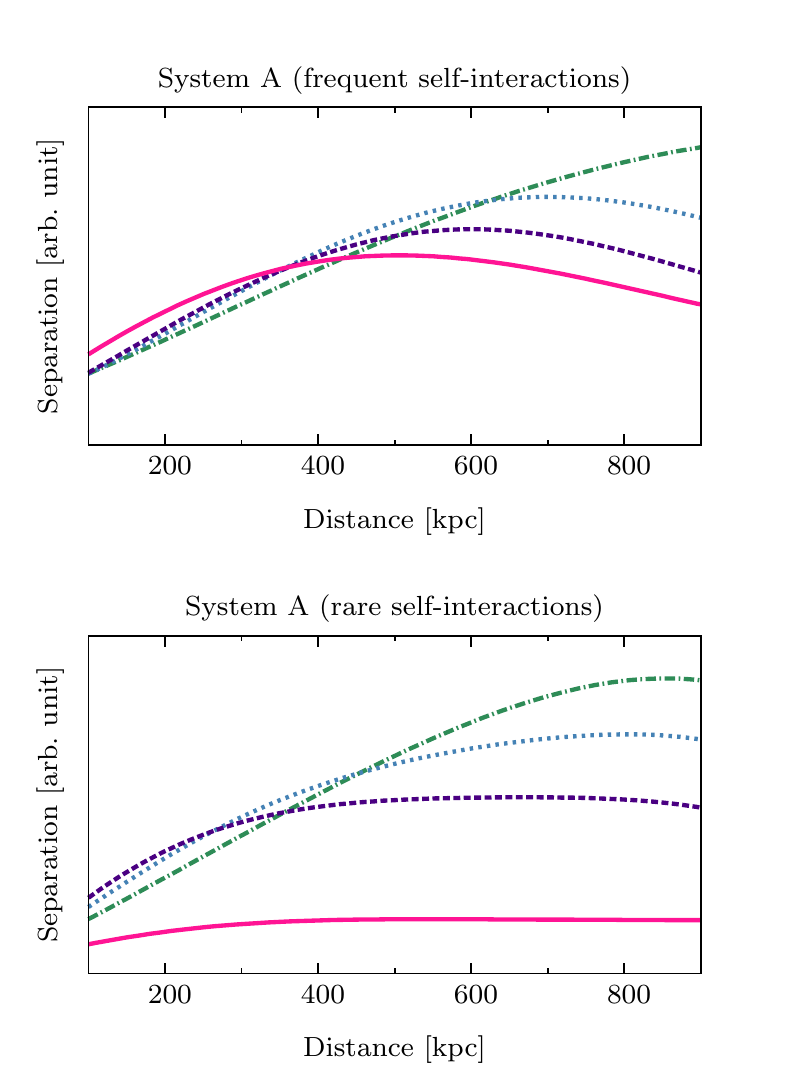}
\vspace{-1mm}
\caption{\label{fig:separations}The expected separation between DM halo and galaxies as a function of the distance between sub-cluster and main cluster for different definitions of the two respective populations. The green dot-dashed line indicates the separation that results when all particles initially bound to the sub-cluster are included in the centroid calculation. For the blue dotted (purple dashed) lines only the inner $80\%$ ($68\%$) of the two distributions are included. For the solid magenta line, we additionally exclude those regions where the surface density of the main cluster dominates. Note the important differences between the case of frequent self-interactions (top panel) and rare self-interactions (bottom panel). System A is representative of Abell 520.}
\end{center}
\end{figure}

By the separation between a DM halo and the corresponding galaxies we mean the difference between the centroids of the two populations, denoted by $\Delta z$. While this definition seems straight-forward, it turns out that our conclusions depend sensitively on just which particles are considered to be part of the respective populations when calculating the centroid positions. We illustrate this problem in Fig.~\ref{fig:separations}, where we show how the separation evolves as a function of the distance between two clusters after they collide for both frequent (top) and rare (bottom) self-interactions. We use a data set generated by our numerical simulation (to be discussed in detail in \S~\ref{sec:freqint}, \S~\ref{sec:rare} and Appendix~\ref{ap:numerical}).

A simple-minded approach would be to just calculate the centroid of the DM halo including \emph{all} DM particles which were initially (i.e.~at infinite negative distance) bound to the sub-cluster. The resulting value would however be strongly biased by particles that have escaped from the DM halo during the collision and are now far away from the peak of the distribution. As a consequence, the separation between the DM halo and galaxies would grow very large as the system evolves in time (green dot-dashed lines in Fig.~\ref{fig:separations}).

For a more refined treatment one should not include particles far away from the peak of the distribution in the calculation of the centroid, since such particles are no longer associated with the DM halo. An easy way to implement this requirement in our simulations would be to select only particles close to the sub-cluster, e.g.~within its tidal radius or scale radius. This approach, however, would not correspond to an observable quantity, because gravitational (weak) lensing is sensitive only to the projected surface density of DM particles. To obtain a realistic estimate of the observable position of the DM halo, which is insensitive to DM particles at large distance, we want to include only regions with a sufficiently large projected density. The blue dotted (purple dashed) lines in Fig.~\ref{fig:separations} indicate the centroid of all DM particles within the iso-density contour containing $80\%$ ($68\%$) of the total mass of the original DM halo. It is clear that a more restrictive choice leads to a smaller separation between the DM halo and galaxies.

We have still neglected an important complication: DM particles that escape from the sub-cluster will typically still be bound to the main cluster and will fall towards its centre. As a result, there will be a relatively high surface density of DM particles originating from the sub-cluster in this region. Nevertheless, we do not want to include these particles in the calculation of the centroid, because they would be associated \emph{observationally} with the main cluster rather than with the sub-cluster. More generally, we only want to include those regions in the centroid calculation where the surface density of DM particles originating from the sub-cluster is \emph{larger} than the background surface density (i.e.~the surface density of the main cluster). The solid magenta line in Fig.~\ref{fig:separations} indicates the resulting separation if we reject regions with large background surface density and then select the central $68\%$ of the remaining DM particles. 

We observe that this definition leads to a \emph{much} smaller separation which does \emph{not} grow with the distance between the two clusters. Moreover, we see that the background subtraction has a much larger effect for rare than for frequent self-interactions~-- a first indication of the differences between the two cases. We will discuss the origin and implications of this difference in \S~\ref{sec:freqint} and \S~\ref{sec:rare}. All plots from now onwards will show the separation based on the central 68\% of the remaining mass after background rejection.

We use the same definition for the position of the galaxies as for the position of the DM halo. We do not consider observational effects which can lead to an apparent separation between DM halo and galaxies
even when there are \emph{no} self-interactions, e.g.~due to systematic biases in the weak lensing reconstruction~\citep{Dawson}. Such considerations may well be important but are beyond the scope of this work.

\bigskip

Before we discuss self-interactions in detail, let us discuss whether a separation between a DM halo and galaxies can arise from gravitational interactions alone. Na\"{\i}vely, it would seem that DM particles passing through the halo are more likely to pass on their energy to similar mass DM particles, rather than to stars, which are much more massive. However, the cross-section for gravitational interactions with large momentum transfer is extremely small. The typical change of energy due to gravitational interactions of a DM particle as it crosses a system is approximately~\citep{Binney2008}
\begin{equation}
\frac{\Delta v^2}{v^2} \approx \frac{8 \log N}{N} \; ,
\end{equation}
where $N$ is the number of DM particles in the system under consideration, typically $\gg 10^{60}$. In other words, the relaxation time for DM particles is so large that we can completely neglect gravitational interactions between individual DM particles and treat the gravitational potential as smooth.

A smooth gravitational potential should affect DM and galaxies in the \emph{same} way and not lead to a separation between the two, provided the initial distribution of DM and galaxies are the same. In many systems, however, the DM halo has a larger spatial extent and might therefore be more susceptible to a change in the external gravitational potential. Moreover, DM particles in the outer parts of the DM halo are more loosely bound and therefore more likely to be tidally stripped. As long as the distance between $S_1$ and $S_2$ is large compared to their respective size, tidal forces lead to symmetric streamers, which do not shift the centroid of the DM distribution. In a cluster collision, on the other hand, tidal forces and dynamical friction may lead to an asymmetric tail of DM particles and therefore potentially a non-zero separation between the DM halo and galaxies even in the absence of DM self-interactions.\footnote{We thank Liliya Williams for raising this point.}

In the present work, we will focus on understanding the separation between DM haloes and galaxies induced by DM self-interactions. To isolate this effect, we consider the case where galaxies and DM particles have a comparable distribution and no separation is expected in the absence of DM self-interactions. Once a separation is confirmed observationally, additional work will be required to understand whether it can be explained in terms of gravitational interactions alone.

\section{Frequent interactions}
\label{sec:freqint}

If we want each DM particle to have a large number of collisions, observations require that the vast majority of these collisions must have very small momentum transfer. Such a large number of collisions can then lead to two observable effects: cumulative evaporation of DM haloes and deceleration of DM haloes. The latter effect is of particular interest in the context of a separation between DM halo and galaxies. Defining the rate of cumulative evaporation as $R_\text{cml} \equiv \dot{E} / E \approx v_0^{-2} \, \mathrm{d}(v^2) / \mathrm{d}t$ and the deceleration rate as $R_\text{dec} \equiv v_0^{-1} \, \mathrm{d} v_\parallel / \mathrm{d}t$, we find (see Appendix~\ref{ap:evaporation})
\begin{equation}
R_\text{cml} \approx R_\text{dec} = \frac{\rho_2 \, v_0 \, \sigma_\mathrm{T}}{2 \, m_\text{DM}} \; ,
\end{equation}
where
\begin{equation}
\sigma_\text{T} = 4 \pi \int_0^{1} \mathrm{d}\cos\theta_\text{cms} \, (1 - \cos \theta_\text{cms}) \, \frac{\mathrm{d}\sigma}{\mathrm{d}\Omega_\text{cms}} \; ,
\end{equation}
is the momentum transfer cross-section. Note that we have restricted the range of integration for $\cos \theta_\text{cms}$ to $\left[0, 1\right]$ and included an additional factor of 2 to account for the fact that the two DM particles are indistinguishable, which has often been neglected in the treatment of DM self-interactions (see Appendix~\ref{ap:evaporation} for a more detailed discussion). We emphasise that these equations are only valid for frequent self-interactions, i.e.~when averaged over a large number of collisions. 

We can see from these equations that cumulative evaporation and deceleration are directly linked~-- it is \emph{impossible} to have deceleration without some cumulative evaporation. Moreover, in some cases the requirement that expulsive collisions are rare ($f \ll 1$) is insufficient to ensure that immediate evaporation will be small compared to cumulative evaporation. In any given model, we therefore need to make sure that $R_\text{imd}$ is comparable to (or smaller than) $R_\text{cml}$. In Appendix~\ref{ap:frequent}, we consider two particular examples of frequent self-interactions, namely long-range interactions and velocity-independent interactions, and confirm in both cases that the fraction of expulsive collisions $f$ as well as the rate of immediate evaporation is sufficiently small. In each case, we find that frequent DM self-interactions lead to a drag force of the form
\begin{equation}
\frac{F_\text{drag}}{m_\text{DM}} = \frac{\tilde{\sigma}}{4\,m_\text{DM}} \rho \, v_0^{2m} \; ,
\end{equation}
where $\tilde{\sigma} = 2 \, \sigma_\mathrm{T}$ in analogy to isotropic scattering (see Appendix~\ref{ap:frequent}).
The case $m=-1$ corresponds to long-range interactions, while the case $m=1$ results from velocity-independent interactions and resembles the drag force that results from ram pressure.

The magnitude of the effective drag force is constrained by observational bounds (e.g.\ on the evaporation rate). The crucial question now is whether observationally allowed drag forces are large enough to lead to a separation between DM haloes and galaxies. Clearly the case of long-range interactions (velocity-independent interactions) is most constrained by systems with small (large) velocities due to the velocity dependence of the evaporation rate. In Appendix~\ref{ap:frequent} we study constraints resulting from dwarf spheroidal galaxies (in the case of long-range interactions) and the Bullet Cluster (in the case of a velocity-independent interaction). We find
\begin{align}
 \frac{\tilde{\sigma}}{m_\text{DM}} & \la 10^{-11}\,\text{cm}^2\,\text{g}^{-1}  \quad (m = -1) \; , \nonumber \\ 
 \frac{\tilde{\sigma}}{m_\text{DM}} & \la 1.2 
\,\text{cm}^2\,\text{g}^{-1}  \quad\quad (m=1) \;.
\label{eq:bounds}
\end{align}
The bound on long-range interactions can be significantly relaxed if the mediator of the interaction has a small but non-zero mass so that DM self-interactions become velocity-independent in systems with low velocity such as dwarf spheroidal galaxies~\citep{Buckley:2009in,Loeb:2010gj,Tulin:2013teo}. Nevertheless, the resulting bounds are still strong enough to ensure that velocity-dependent self-interactions cannot have any observable effect on cluster collisions, which have typical velocities of $1000\,\text{km\,s}^{-1}$ or more. In the following we will therefore consider only velocity-independent interactions in the context of galaxy clusters.

Let us first discuss qualitatively under which conditions a separation between DM halo and stars/galaxies can arise and then confirm our expectations with explicit numerical simulations.

\subsection{Analytical arguments}

A drag force due to DM self-interactions will primarily affect all the DM particles in the halo but not the galaxies. Consequently, in the (decelerating) frame of the DM halo, the galaxies will experience a fictitious (accelerating) force. If we treat this acceleration as approximately constant, it can be modelled as arising from a linear potential. Combining this with the gravitational potential of the DM halo, we obtain an effective potential that describes the motion of galaxies in the frame of the halo:
\begin{equation}
\Phi_\text{eff}(\bmath{x}) = \Phi_\text{g}(\bmath{x}) + \frac{\tilde{\sigma}}{4\,m_\text{DM}} \rho \, v_0^{2m-1} \, \bmath{x} \cdot \bmath{v}_0
\end{equation}
Clearly, because of the tilt of the potential, not all galaxies will be bound. Those that remain bound will also remain close to the position of the DM halo but those that are no longer bound will end up travelling ahead of the halo, thus leading to an apparent separation. In the presence of a drag force, the DM halo will therefore retain its shape, while some fraction of the galaxies will travel ahead of the halo.

\begin{table}
\caption{Summary of the parameters used in the numerical simulations. All clusters are modelled using Hernquist profiles. See Appendix~\ref{ap:numerical} for details. System A is representative of Abell 520, System B of the Bullet Cluster and System C of the Musket Ball Cluster.}
\label{tab:parameters}
\centering
\begin{tabular}{@{}llccc}
\hline
& & System A & System B & System C \\
\hline
\multicolumn{2}{l}{\hspace{-2mm}Sub-cluster} & & & \\
$M$ & $\left[ \, 10^{14}\,\mathrm{M}_{\sun} \, \right]$ & 1.5 & 3.0 & 1.5 \\
$b$ & $\left[ \, \text{kpc} \, \right]$ & 200 & 100 & 300 \\
\hline
\multicolumn{2}{l}{\hspace{-2mm}Central cluster} & & & \\
$M$ & $\left[ \, 10^{14}\,\mathrm{M}_{\sun} \, \right]$ & 3.5 & 25 & 3.0 \\
$b$ & $\left[ \, \text{kpc} \, \right]$ & 400 & 1000 & 400 \\
$v_\text{dis}$ & $\left[ \, \text{km\,s}^{-1} \, \right]$ & 1000 & 1200 & 800 \\
\hline
\multicolumn{2}{l}{\hspace{-2mm}Collision} & & & \\
$v_\text{col}$ & $\left[ \, \text{km\,s}^{-1} \, \right]$ & 2400 & 4500 & 2000 \\
\hline
\end{tabular}
\end{table}

Let us estimate the potential magnitude of the effects of a drag force in cluster collisions. For simplicity we assume that a galaxy will leave the sub-cluster if the deceleration of the halo exceeds the gravitational acceleration of the galaxy. For the sub-cluster in the Bullet Cluster we adopt the parameterisation given in Table~\ref{tab:parameters}, leading to
\begin{equation}
a_\text{grav}(r) = - 4.2 \times 10^{-9}\,\text{m\,s}^{-2} \left(1 + \frac{r}{100\,\text{kpc}}\right)^{-2} \; .
\end{equation}
On the other hand, the observational bounds in Eq.~(\ref{eq:bounds}) require $a_\text{drag} \la 10^{-9}\,\text{m s}^{-2}$. Saturating this bound, we find that $a_\text{grav} < a_\text{drag}$ for all galaxies at a distance $r > r_s$ from the centre of the Bullet, corresponding to $75\%$ of the total number.\footnote{As discussed in \S~\ref{sec:separation} we assume that DM particles and galaxies have the same phase space distribution.} We conclude that a drag force can have potentially \emph{large} effects in colliding galaxy clusters. For more quantitative estimates, we will rely on a numerical simulation, which we introduce next.

\subsection{Numerical simulation}

As discussed above, a separation between a DM halo and the galaxies initially bound to it can arise if the halo experiences a drag force comparable to the gravitational acceleration of particles within the system. To obtain a quantitative estimate of this effect, we perform a numerical simulation. Rather than use a computationally expensive $N$-body simulation where the gravitational forces at every point in time and space are calculated from the simulated particles, we simply trace the motion of a set of test particles in a (time-dependent) gravitational potential. In particular we consider three different systems: System A is representative of Abell 520~\citep{Mahdavi:2007yp, Jee:2012sr}, System B of the Bullet Cluster~\citep{Markevitch:2003at,Clowe:2006eq} and System C of the Musket Ball Cluster~\citep{2012ApJ...747L..42D, Dawson:2012fx}. For details of how we model these clusters see Table~\ref{tab:parameters} and Appendix~\ref{ap:numerical}. A discussion of other interesting systems is included at the end of this section.

\begin{figure}
\begin{center}
\includegraphics[width=0.8 \columnwidth]{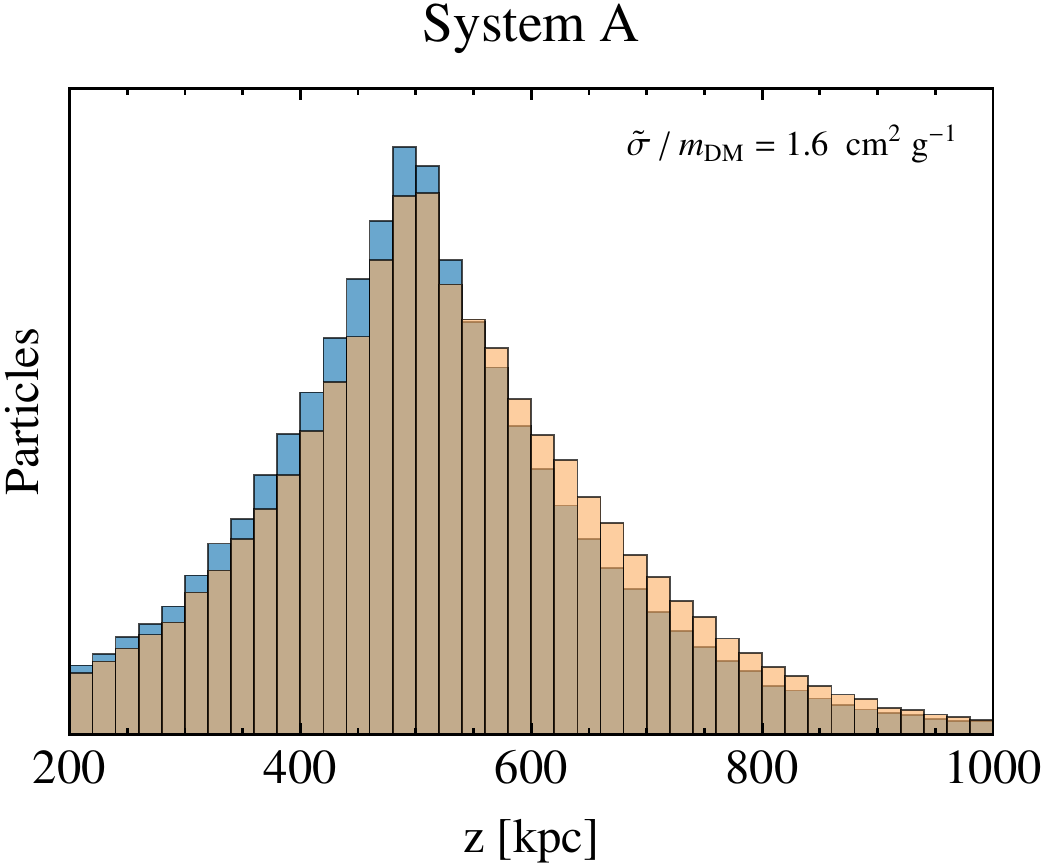}
\end{center}
\vspace{1mm}
\begin{center}
\includegraphics[width=0.8 \columnwidth]{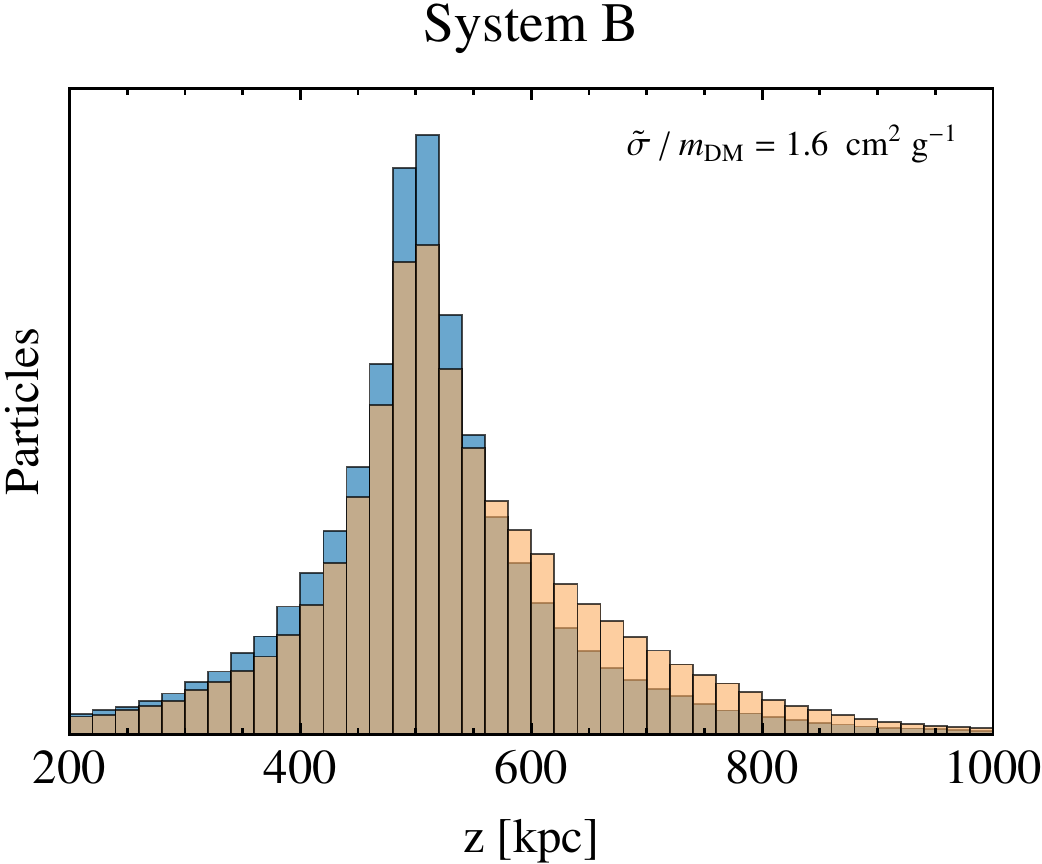}
\vspace{1mm}
\caption{\label{fig:drag_histo}One-dimensional distribution of DM particles (blue) and galaxies (orange) along the line of motion of the DM sub-cluster for System A (representative of Abell 520) and System B (representative of the Bullet Cluster) for frequent self-interactions. The relative normalisation of the two distributions has been chosen in such a way as to facilitate comparison. One can clearly distinguish the tail of galaxies which are moving ahead of the DM halo.}
\end{center}
\end{figure}

Details of our simulation are given in Appendix~\ref{ap:numerical}. We run the simulation with $N = 2 \times 10^5$ particles for five different choices of $\tilde{\sigma} / m_\text{DM}$ ranging between 0 and $1.6\,\text{cm}^{2}\,\text{g}^{-1}$ and show the results in Fig.~\ref{fig:drag_histo}. In agreement with our expectations, we observe that the DM halo retains its shape throughout the simulation (apart from a small fraction of DM particles, which are stripped from the DM halo by tidal forces). The distribution of galaxies, on the other hand, becomes asymmetric and develops a tail in the forward direction due to the asymmetry in the effective potential as well as galaxies escaping from the halo.

\begin{figure*}
\begin{center}
\includegraphics{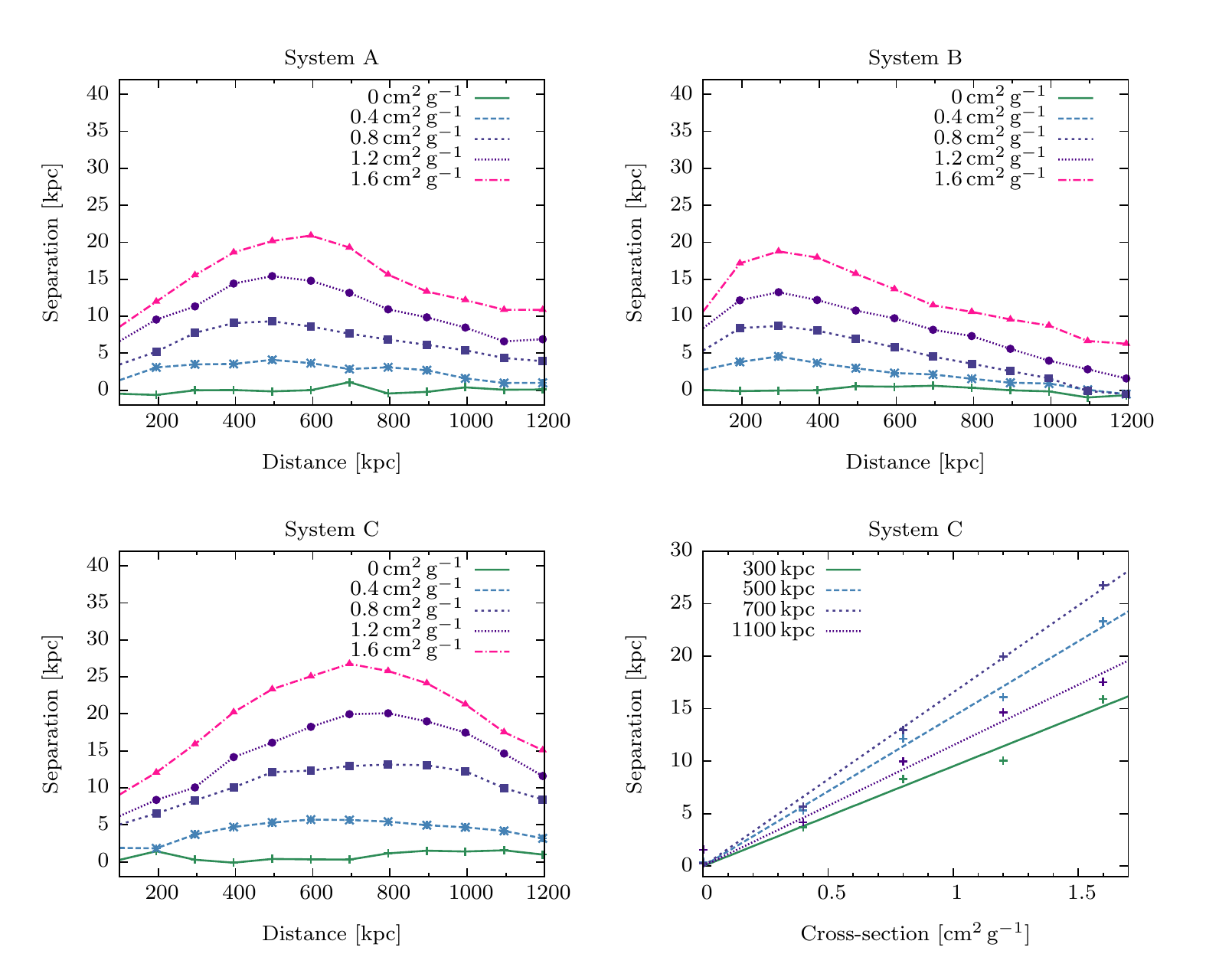}
\vspace{-2.5mm}
\caption{\label{fig:drag_separation}Observed separation resulting from frequent DM self-interactions for various DM self-interaction cross-sections as a function of the cluster distance for System A (representative of Abell 520), System B (representative of the Bullet Cluster) and System C (representative of the Musket Ball Cluster). The bottom-right panel also shows for System C the separation as a function of the cross-section for various distances. Note that a self-interaction cross-section of $\tilde{\sigma} / m_\text{DM} = 1.6\,\text{cm}^{2}\,\text{g}^{-1}$ implies an evaporation of up to $40\%$ of the total sub-cluster mass in the Bullet Cluster, in slight tension with observations.}
\end{center}
\end{figure*}

In order to obtain a quantitative estimate of the separation between DM halo and galaxies, we calculate the centroids of the two respective distributions as described in \S~\ref{sec:separation}. We present our results in Fig.~\ref{fig:drag_separation}. As expected, we find for all systems that the separation between DM halo and galaxies increases with increasing drag force (see bottom-right panel of Fig.~\ref{fig:drag_separation}). Furthermore, we observe that the separation grows initially as the sub-cluster moves away from the main cluster. This agrees with our expectations as the unbound galaxies move ahead of the DM halo with a slightly larger velocity and the bound galaxies require some time to reach their maximum displacement. With decreasing drag force at larger distances, however, the separation also decreases. The reason is that the galaxies that escaped from the halo are no longer considered part of the system and are therefore eventually excluded from the calculation of the centroids, while the galaxies that remained bound return to their original position.

An important observation is that the maximum separation in System B is smaller than in Systems A and C and also decreases more quickly after the collision, even though the main cluster is more massive and more extended in System B. However, the larger background density also implies that the rejection of galaxies at large distances is more stringent for System B than for Systems A and C. Consequently, the position of the centroid is dominated by galaxies which remain bound to the DM halo throughout the collision. Since the sub-cluster in System B is very tightly bound, these galaxies will return to their original position very quickly, once the sub-cluster moves away from the region of highest density. In other words, the larger drag force is balanced by a stronger gravitational pull.

Even if we saturate the bounds on the drag force from Eq.~(\ref{eq:bounds}), the predicted separations in all the systems under consideration are smaller than $25\,\text{kpc}$. In particular, the separation in the Bullet Cluster is expected to be below the current observational bound $\Delta z \la 50\,\text{kpc}$~\citep{Randall:2007ph}. Nevertheless, our estimates demonstrate that colliding clusters are potentially sensitive probes of frequent self-interactions and that systems like Abell 520 and the Musket Ball Cluster are interesting to study in spite of the smaller surface densities compared to the Bullet Cluster.

Finally, we would like to emphasise that the results discussed above are somewhat optimistic, because we have assumed the same phase space distribution for DM and galaxies. In a more realistic description galaxies will typically have lost some kinetic energy due to dissipation and will therefore sit deeper in the gravitational potential than DM particles, thus are less likely to escape from the DM halo. Taking this effect into account, the separation between DM haloes and galaxies is expected to be somewhat \emph{smaller} than predicted by our numerical simulations. Nevertheless they may be large enough to be detectable.

\subsection{Other systems of interest}

To conclude the discussion of frequent self-interactions, we consider other systems where effective drag forces may be of interest. As we have discussed above, it is crucial that at some point in the evolution of the system the drag force on the DM halo was comparable to the typical gravitational acceleration, i.e.~that the DM halo has passed through a region of large DM density. The magnitude of the expected separation is therefore not necessarily proportional to the integrated DM density probed by the DM halo and it is not sufficient to have a highly evolved system in order to get a large separation.

\citet{2011MNRAS.415..448W} observed a DM halo inside a galaxy cluster, which seems to be \emph{entirely separated} (by several kpc) from the nearest population of stars (see also \citet{Mohammed}). This has been interpreted as evidence for an effective drag force affecting the DM halo. Given our analytical arguments and numerical results from above, this interpretation seems very unlikely. If the DM halo moves on circular orbits, it will never have probed a DM density high enough to be strongly affected by self-interactions and the gravitational interactions can compensate the drag force at least for the more tightly bound stars. The same reasoning applies if the DM halo is falling towards the centre, but has not yet passed through a region of high DM density. If it has actually passed through the central region of the galaxy cluster, on the other hand, DM self-interactions have to be small in order for the DM halo not to be destroyed, so that again no complete separation between DM halo and stars is expected.

We note, however, that sub-clusters falling into a main cluster may exhibit a partial separation between DM and stars in the presence of a large drag force on the DM halo even if they get destroyed once they reach the central region. It is therefore a very promising route for constraining or measuring frequent self-interactions to analyse galaxy clusters and statistically determine the separation between DM and stars in infalling sub-clusters~\citep{Massey, Harvey:2013bfa, Harvey}.

Finally, we return to the case of long-range interactions, which can be best probed by low velocity systems such as dwarf spheroidal galaxies. We make a crude estimate for a dwarf spheroidal galaxy by assuming an Navarro-Frenk-White profile (NFW profile) with $r_s = 0.1$ kpc and $\rho_0 = 4\times10^7 \, \mathrm{M}_{\sun}\,\text{kpc}^{-3}$. For $r \sim r_s$ the gravitational acceleration is then given by $a_\text{grav}(r) \sim 10^{-12}\,\text{m\,s}^{-2}$, which is comparable to the observational bound on the drag force in Eq.~(\ref{eq:bounds}). For stars with $r > r_s$ the drag force can be significantly larger than the gravitational pull on the stars and therefore potentially enables these stars to escape from the DM halo. Consequently, long-range interactions can have potentially large effects on dwarf spheroidals. We leave a detailed study of these effects to future work and concentrate on cluster collisions in the remainder of this paper.

\section{Rare self-interactions}
\label{sec:rare}

Let us now turn to the case where DM self-interactions are rare but typically have large momentum transfer. We will focus on the case of isotropic scattering. In this case, the fraction of expulsive collisions is large ($f = 60\%$ for the Bullet Cluster), meaning that the cross-section is strongly constrained by bounds on the evaporation rate of a given system. For example, if we require that the sub-cluster in the Bullet Cluster loses no more than $30\%$ of its mass during the collision with the main cluster~\citep{Markevitch:2003at}, we obtain the constraint
\begin{equation}
\frac{\Sigma_2 \, f \, \sigma }{m_\text{DM}} \la 0.3 \; .
\end{equation} 
Taking $\Sigma_2 = 0.3\, \text{g\,cm}^{-2}$~\citep{Markevitch:2003at}, we find
\begin{equation}
\frac{\sigma}{m_\text{DM}} \la 1.7\, \text{cm}^2\,\text{g}^{-1} \; .
\end{equation}
By construction, such a value of the cross-section implies that a significant fraction of the DM particles in the sub-cluster will pass through the main cluster \emph{without} a single collision. More generally, for a DM particle travelling through a DM halo of density $0.1\,\text{GeV\,cm}^{-3}$ such a cross-section corresponds to a mean free path of around $1\,\text{Mpc}$. In a typical DM halo, a DM particle will therefore complete many orbits without scattering.

\begin{figure}
\begin{center}
\includegraphics[width=0.9 \columnwidth]{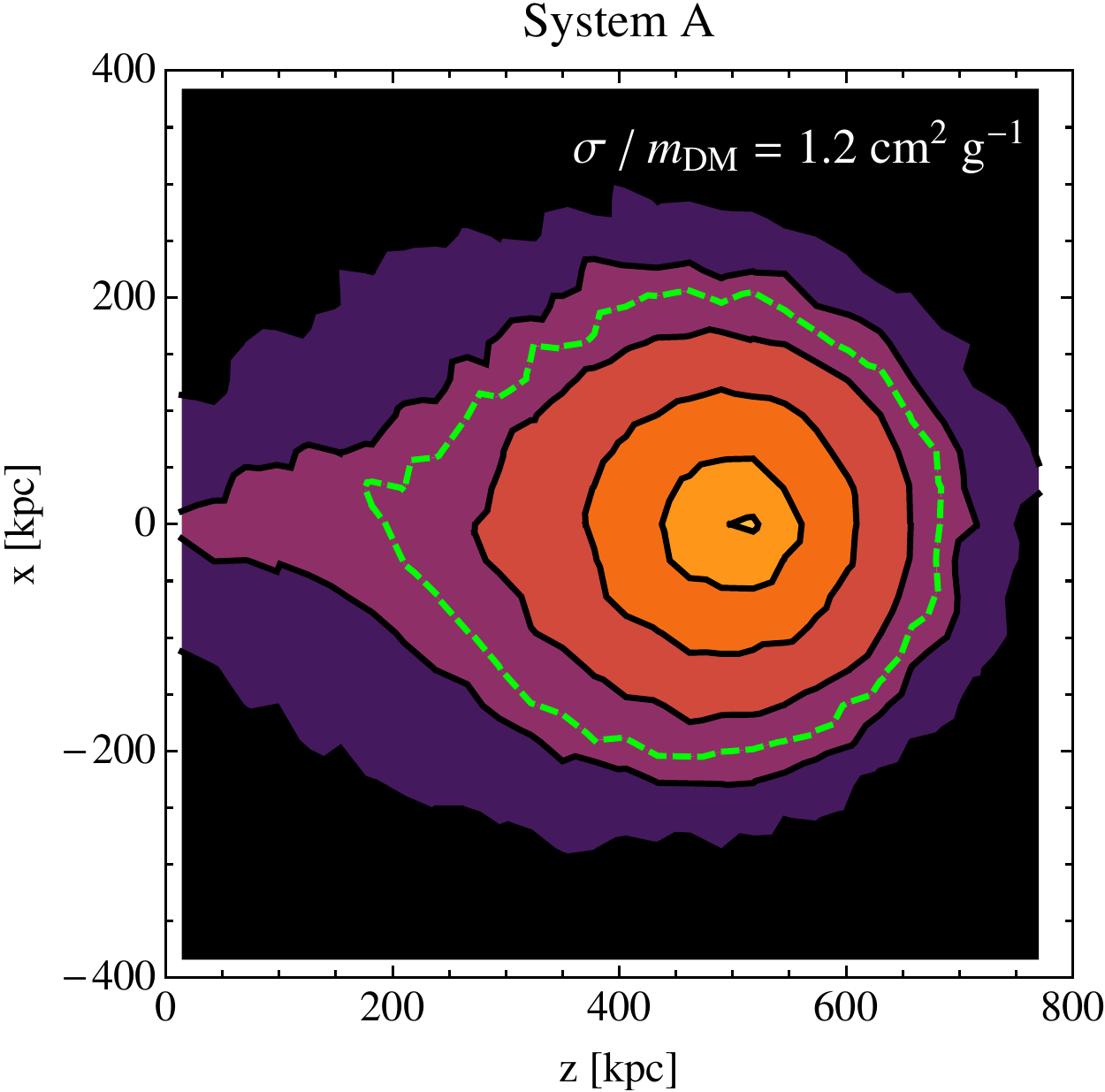}
\end{center}
\vspace{1mm}
\begin{center}
\includegraphics[width=0.9 \columnwidth]{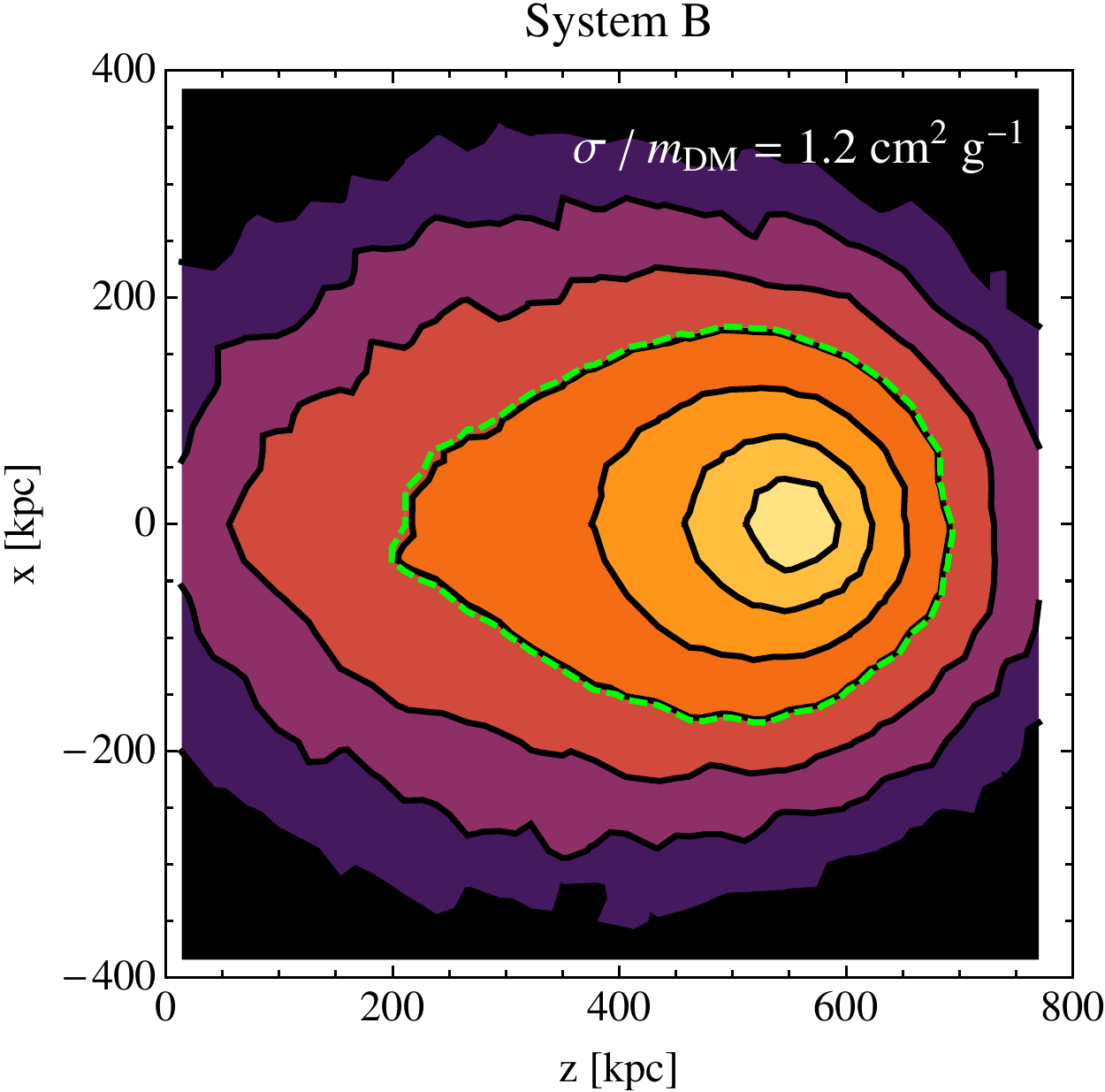}
\vspace{1mm}
\caption{\label{fig:2D}Two-dimensional distribution (i.e.~surface density) of DM after the sub-cluster has passed through the main cluster for System A (representative of Abell 520) and System B (representative of the Bullet Cluster) for the case of contact interactions. The black solid contours indicate lines of constant surface density, starting at $10^8\,\mathrm{M}_{\sun}\,\text{kpc}^{-2}$ at the outermost contour and increasing by a factor of 2 with each contour towards the centre. For example, the dark purple region (outermost for both systems) has a surface density of $\Sigma = (\text{1--2}) \times 10^8\,\mathrm{M}_{\sun}\,\text{kpc}^{-2}$, the light yellow region (innermost for System B) has a surface density of $\Sigma > 6.4 \times 10^9\,\mathrm{M}_{\sun}\,\text{kpc}^{-2}$. The green dashed contour indicates the iso-density contour containing $68\%$ of the total halo mass, which is used for the calculation of the centroid (see \S~\ref{sec:separation}).}
\end{center}
\end{figure}

The crucial question now is how DM particles that do not directly experience any collisions will be affected by the scattering of DM particles in their vicinity. For a system moving through a background DM density at a high velocity, DM particles will typically scatter in the direction \emph{opposite} to the direction of motion, and a large fraction of the scattered particles will no longer be bound to the DM halo. As they leave the system, they will slow down in the gravitational potential of the DM halo thus transferring some of their momentum to the surroundings. In other words, the tail of scattered DM particles will exert a gravitational pull on the DM halo, which will slow down the entire system.

One might be tempted to conclude that rare self-interactions therefore lead to a drag on the DM halo similar to the one we found for the case of frequent self-interactions. However, the origin of this drag are \emph{gravitational interactions alone}. Consequently, this drag force will necessarily affect the DM halo and the galaxies and stars within it in exactly the same way.\footnote{As discussed in \S~\ref{sec:separation} scattering between individual DM particles with large momentum transfer via gravitational interactions is completely negligible.} Our central observation is therefore that a DM particle that does not directly experience any collisions will behave exactly like a collisionless galaxy.

A possible exception would be if a DM particle, after having scattered, scatters again as it leaves the DM halo. If such secondary scatterings were to occur frequently, DM particles would transfer their momentum preferentially to the DM halo rather than to stars and galaxies. However, observational constraints on evaporation rates and, in fact, halo shapes imply that the probability for a DM particle to scatter within one orbit has to be very small. In other words, most particles that scatter from DM particles in the main cluster will typically not scatter again as they leave the sub-cluster.

We conclude that rare DM self-interactions do \emph{not} lead to an effective drag force that can separate DM halo and galaxies. DM particles which have not undergone any collisions will always remain coincident with the equally collisionless galaxies. However, those particles which have had collisions will preferentially travel towards the back of the halo. Ultimately, these particles will end up far away from their original system, but shortly after the collision they still appear to be a part of the DM halo. Consequently, as they leave the system, these particles will shift the centroid of mass of the DM halo in the direction opposite to the direction of motion thus leading to an \emph{apparent} separation between DM and galaxies shortly after the collision.

\begin{figure}
\begin{center}
\includegraphics[width=0.8 \columnwidth]{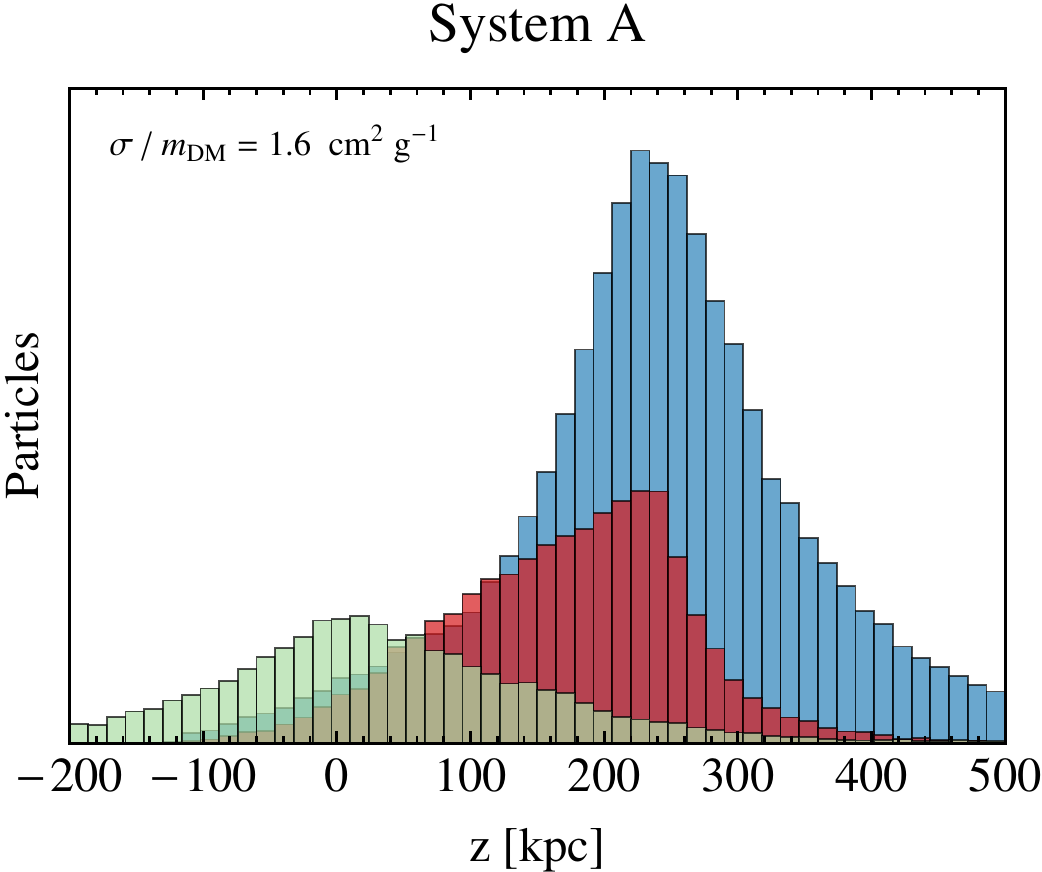}
\end{center}
\vspace{1mm}
\begin{center}
\includegraphics[width=0.8 \columnwidth]{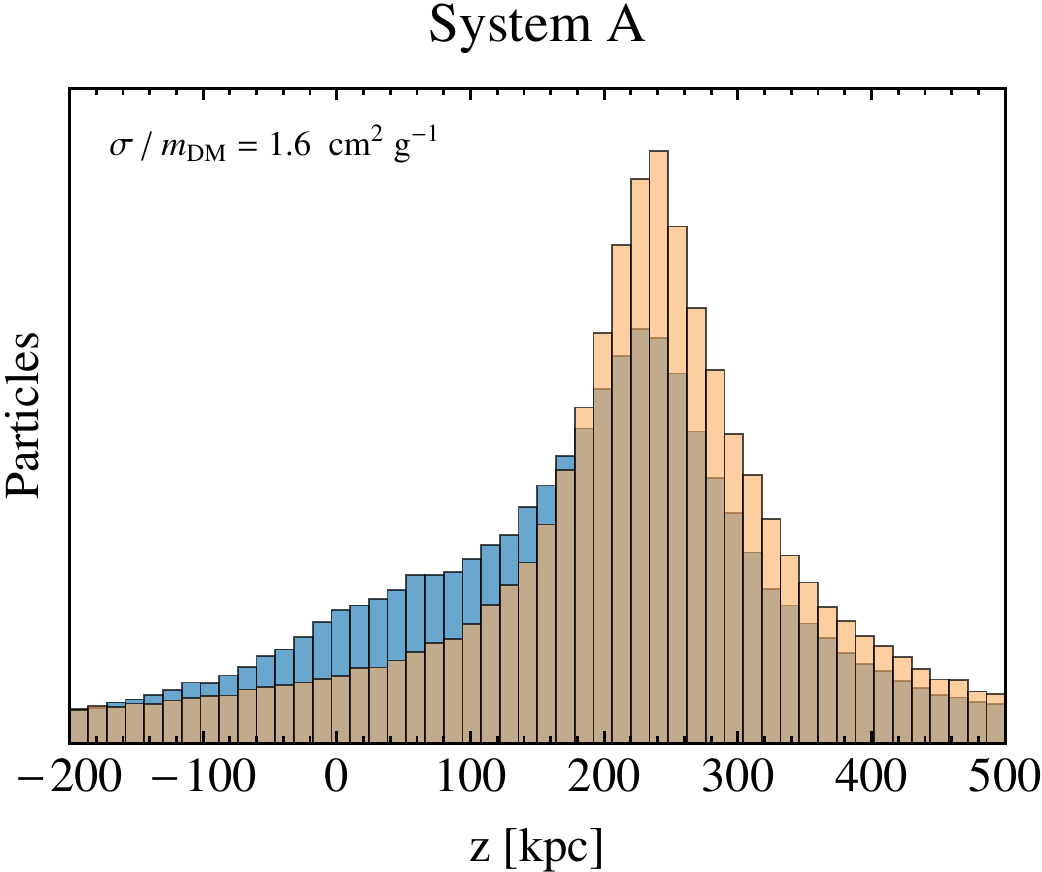}
\vspace{1mm}
\caption{\label{fig:Histo}One-dimensional distribution of simulated DM particles and galaxies along the direction of motion of the sub-cluster for the case of contact interactions. The top panel shows three different populations of DM  particles: Particles which have scattered over the course of the simulation depending on whether they are still bound (red) or not bound (green) to the sub-cluster and particles which have not scattered during the simulation (blue). The bottom panel shows the sum of these contributions (blue) compared to the distribution of galaxies (orange). The relative normalisation of these two distributions has been chosen in such a way as to facilitate comparison. Note that System A is representative of Abell 520.}
\end{center}
\end{figure}

A similar argument applies to particles that have scattered but remain bound to the DM halo. These particles will typically have elliptical orbits. Since the relaxation time for DM particles is very large, we expect them to retain these orbits for a long time. For a short time after the collision (i.e.~before they complete half an orbit) these particles will therefore preferentially be found towards the back of the system. Again, particles that have scattered very recently can induce an apparent separation between DM and galaxies.

We have identified a \emph{key difference} between rare and frequent DM self-interactions. For rare self-interactions, a separation between DM halo and galaxies is caused by DM particles leaving the gravitational potential in the direction opposite to the direction of motion. This is in contrast to the case of frequent self-interactions, where the separation arises from galaxies moving \emph{ahead} and leaving the gravitational potential in the direction of motion. Consequently, the two scenarios are distinguishable if the shape of the stellar distribution can be measured with sufficient accuracy (and the initial distribution is known).

To check our expectations, we have extended the numerical simulation introduced above to include contact interactions between individual DM particles. The details of our code are presented in Appendix~\ref{ap:numerical}. Fig.~\ref{fig:2D} shows the resulting shapes of the DM haloes after the cluster collision for two different systems. As expected, most DM particles are still bound to the sub-cluster even long after the collision, but the asymmetry of the iso-density contours resulting from the tail of scattered DM particles can be clearly seen. 

Of the DM particles which have scattered, some will remain bound to the sub-cluster, while others will escape from the system. Since our simulation is based on a smooth gravitational potential, it is easy to determine whether a DM particle is bound to the sub-cluster or not. We can therefore plot these two populations separately, as shown in the top panel of Fig.~\ref{fig:Histo}. As expected, particles which have received such a large momentum transfer that they are no longer bound to the sub-cluster are found furthest from the sub-cluster. On the other hand, those particles which have scattered but remained bound are only slightly shifted relative to the particles which have not scattered.

\begin{figure*}
\begin{center}
\includegraphics{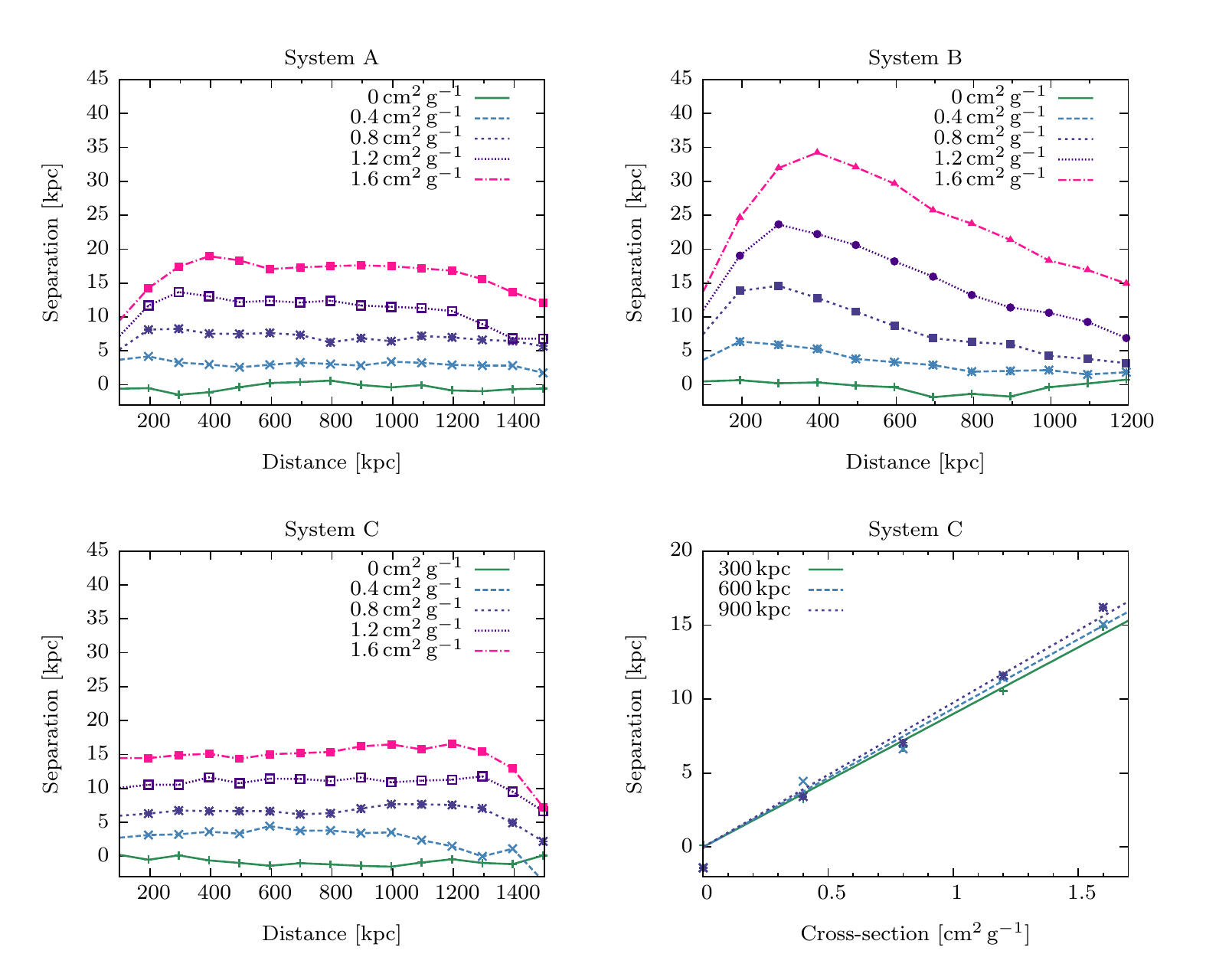}
\vspace{-2.5mm}
\caption{\label{fig:results} Observed separation resulting from rare DM self-interactions for various DM self-interaction cross-sections as a function of the cluster distance for System A (representative of Abell 520), System B (representative of the Bullet Cluster) and System C (representative of the Musket Ball Cluster). The bottom-right panel also shows for System C the separation as a function of the cross-section for various distances.}
\end{center}
\end{figure*}

In the bottom panel of Fig.~\ref{fig:Histo} we compare the total distribution of DM particles with the total distribution of galaxies. It is obvious from this plot that the peaks of the two distributions are perfectly coincident~-- in agreement with our expectation. However, it is also clear that the tails of the distributions differ for the two populations, leading to a separation of their centroids. We can calculate this separation in the same way as for the case of frequent self-interactions.

Our results are shown in Fig.~\ref{fig:results} for Systems A, B (top row) and C (bottom row) as introduced in Appendix~\ref{ap:numerical} (see also Table~\ref{tab:parameters}). We find separations between the centroid of the DM halo and the centroid of the galaxies which can be large shortly after the collision, but then decrease with time. The typical magnitudes which we observe (10--30 kpc) are comparable to separations found for similar cross-sections by~\citet{Randall:2007ph}.\footnote{Note that our definition of the separation~-- in particular the background rejection~-- is very conservative and larger separations are expected for more optimistic choices (cf.~Fig.~\ref{fig:separations}). An accurate comparison with the values from~\citet{Randall:2007ph} would require knowledge of exactly how the authors reconstruct the DM halo and subtract backgrounds.} We also confirm the observation that the separation is approximately proportional to $\sigma / m_\text{DM}$ (see bottom-right panel of Fig.~\ref{fig:results}).

An important feature, which can be inferred from Fig.~\ref{fig:results}, is that in different systems the separation evolves differently with time (or distance between the two clusters). Systems A and C show a similar behaviour: The separation grows quickly after the collision of the two clusters, peaks at a distance of a few hundred kpc and then decreases very slowly as the clusters move further apart. In System B, on the other hand, the separation decreases much more quickly. The reason for this difference is the large asymmetry between main cluster and sub-cluster in System B. The background density of DM particles from the main cluster is so large, that DM particles leaving the sub-cluster will very quickly become indistinguishable from the main cluster. Consequently, particles which escape from the sub-cluster can only contribute to the separation for a very short time. At later times, the separation arises only from particles that have remained bound to the sub-cluster during the collision. Since the sub-cluster in System B is very tightly bound, these particles will quickly reach their maximum distance and begin to fall back towards to centre of the sub-cluster. As a consequence, the separation decreases and can even become negative at very late times if the scattered particles overtake the collisionless matter.

\begin{figure}
\begin{center}
\includegraphics[width=0.8 \columnwidth]{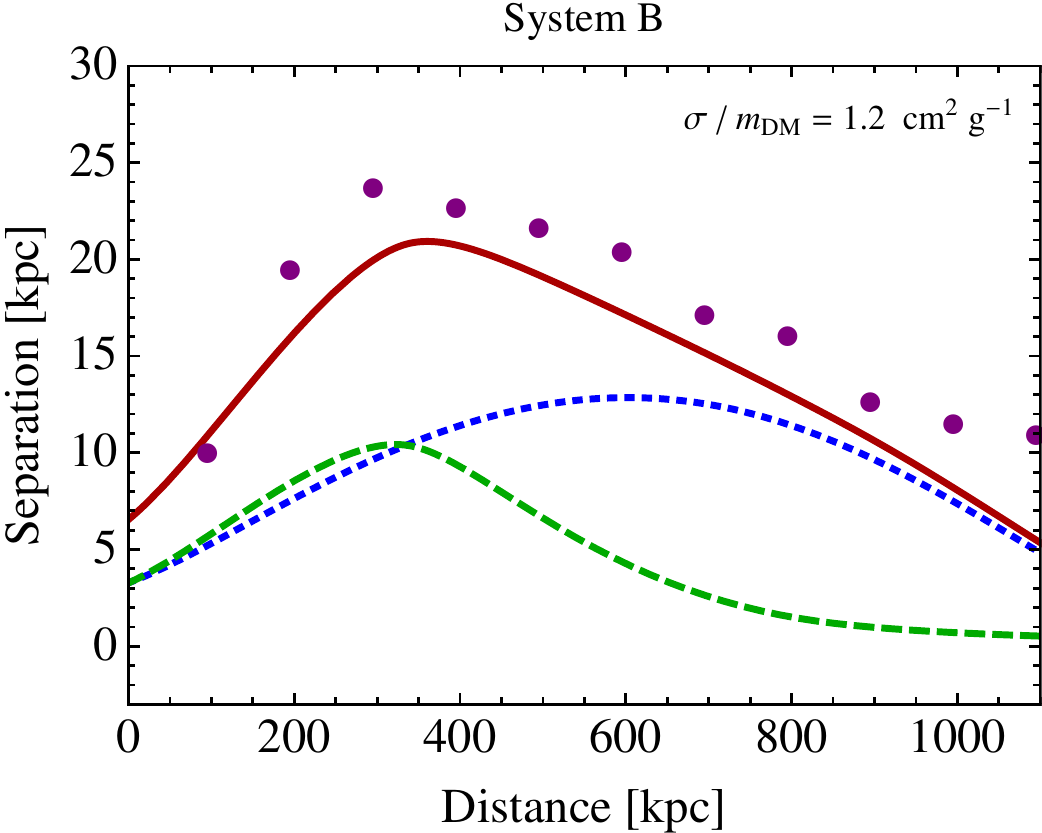}
\vspace{1mm}
\caption{\label{fig:analytical}Observed separation in System B as a function of the distance between main cluster and sub-cluster for rare DM self-interactions with $\sigma/m_\text{DM} = 1.2\,\text{cm}^2\,\text{g}^{-1}$ according to our analytical model (solid red line) and our numerical simulation (purple dots). The dashed green (dotted blue) line indicates the contribution to the total separation from particles with $v > v_\text{esc}$ ($v < v_\text{esc}$).}
\end{center}
\end{figure}

In Appendix~\ref{ap:analytical} we introduce a simple analytical model of rare self-interactions based on tracing the orbits of individual DM particles after a collision. The separation predicted by this model for System B is shown in Fig.~\ref{fig:analytical} (solid red line). We find good agreement with the separation obtained by our numerical simulation (purple dots). As in the numerical simulation, we can make explicit the contribution arising from particles with $v > v_\text{esc}$, which escape from the sub-cluster and particles with $v < v_\text{esc}$, which remain bound to the sub-cluster after scattering and travel on elliptical orbits (cf.~Fig.~\ref{fig:Histo}). The two separate contributions are shown in Fig.~\ref{fig:analytical} as green (dashed) and blue (dotted) lines, respectively. It is clearly seen that at small distances between the two clusters, the separation is dominated by particles with $v > v_\text{esc}$, while the dominant contribution at large distances results from particles with $v < v_\text{esc}$. 

Modelling Systems A and C is more complicated, because the background densities are smaller and therefore the tails of the distributions are more important. Moreover, we cannot neglect the velocity dispersion of the sub-cluster compared to the collision velocity $v_0$. As a result, we obtain an additional contribution to the separation from particles that only receive a small momentum transfer but still have sufficient energy to escape from the sub-cluster. In projection, these particles will lead to a separation that slowly grows in time and give the dominant contribution at very large distances, contributing to the differences between Systems A, B and C observed in Fig.~\ref{fig:results}.

In Appendix~\ref{ap:analytical}, we also make an estimate of the typical time and distance scales for the separation. We find that the largest separation is expected for $t \sim t_\text{esc}$, which is the time it takes for particles with $v \sim v_\text{esc}$ to leave the sub-cluster. In terms of the mass $M$ and the size $b$ of the sub-cluster, we find $t_\text{esc} \sim 2 \sqrt{b^3 / (G_\mathrm{N} \, M)}$. Particles that remain bound in a collision will typically take a time $t_\text{orb} \sim 3 \, t_\text{esc}$ to complete half an orbit. For $t > t_\text{orb}$, these particles will give a negligible or even negative contribution to the total separation.

Using our simple analytical model, we expect that the maximum separation is approximately given by
\begin{equation}
\Delta z_\text{max} = \frac{f \, \Sigma_2 \, \sigma \, b}{m_\text{DM}} \; .
\end{equation}
For all three systems in our simulations we find approximately $f \approx 0.6$. The averaged surface density $\Sigma_2$ in System B is larger than the one in System A (C) by a factor of about 3 (6). On the other hand, the size of the sub-cluster $b$ is smaller by a factor of 2 (3). Consequently, we expect the maximum separation to be larger by a factor of 1.5 (2)~-- in good agreement with our numerical simulations. 

An important implication is that the separation in System B is \emph{not} significantly larger than the separation in Systems A and C, even though the surface density of the main cluster is much larger in System B. The reason is that System B (i.e.~the Bullet Cluster) is much smaller and much more tightly bound than the sub-clusters in Systems A and C, and therefore DM particles that have scattered either leave the system very quickly, or remain on orbits relatively close to the centre of the sub-cluster. We conclude that the properties and dynamics of the sub-cluster have an important influence on the magnitude and time-evolution of the expected separation.

\section{Conclusions}

We have studied the effects of DM self-interactions in collisions of galaxy clusters with particular focus on the resulting separation between the DM halo of the sub-cluster and the collisionless galaxies. In our analysis we have made the distinction between frequent collisions with small momentum transfer and rare collisions with large momentum transfer. Only when the fraction of expulsive collisions $f$ is much smaller than unity, is it possible to have frequent DM self-interactions without violating observational constraints on the evaporation rate. When this is the case, DM self-interactions can be described by an effective drag force. However when $f$ is large, DM self-interactions must be rare and an effective description of collective effects is not possible since only some fraction of the DM particles are affected at all. Overall we find significant differences between these two classes of DM self-interactions concerning both qualitative and quantitative predictions. It is therefore important to make a clear distinction between them.

For all types of self-interaction that we have considered, the peak of the DM distribution remains coincident with the peak of the distribution of galaxies, after conservative bounds on the DM self-interaction cross-section have been imposed. In other words, the effect of self-interactions is never large enough to completely separate DM halo and galaxies and both will remain bound to the same gravitational potential. Nevertheless, in the presence of DM self-interactions, the flanks and tails of the two distributions will be deformed in different ways, leading to a separation of the centroids of the two distributions. 

The nature of this deformation depends sensitively on the nature of the DM self-interactions. If self-interactions are rare but involve large momentum transfer, only a small fraction of DM particles will scatter, but those that do will either leave the system in the direction opposite to the direction of motion, or travel on highly elliptical orbits. As a result, the DM distribution develops a tail in the backward direction, while the distribution of galaxies remains unaffected. If, on the other hand, self-interactions are frequent and have typically small momentum transfer (i.e.~a strongly forward-peaked differential cross-section), all DM particles will have a large number of collisions. The resulting deceleration of the entire system will cause loosely bound galaxies to escape from the system and travel ahead. Consequently, the distribution of galaxies will develop a tail in the forward direction, while the distribution of DM retains its shape. 

This difference is in principle detectable with advances in observational astronomy and may be a promising route to distinguish a drag force due to frequent DM self-interactions from rare scattering due to contact interactions. Although assumptions must be made as to what the shape of the distribution was \emph{before} the collision in order to determine the distortion, this approach is potentially a direct way to observationally probe a key property of DM, which would strongly discriminate between the many proposed particle candidates.

To check our expectations based on analytical arguments, we have developed a numerical simulation capable of studying cluster collisions in the presence of either rare or frequent DM self-interactions. The simulation traces the motion of test particles in a smooth gravitational potential and calculates the effects of self-interactions based on the surrounding DM density. This approach is very fast and quite adequate to provide a qualitative understanding of the impact of self-interactions and confirm our expectations, even though it does not fully capture all gravitational effects. A detailed $N$-body simulation will be needed to study these effects in detail as well as include additional effects of DM self-interactions such as the formation of constant density cores in galaxy clusters. However, to our knowledge no consistent implementation of frequent self-interactions into $N$-body simulations has been achieved so far and doing so would require significant further work.

We emphasize that the separation between DM halo and galaxies does not always grow in time, but typically peaks a short time after the collision. The reason is that the tails of the two distributions will not be observable once their (projected) density becomes comparable to the background density. Therefore, the observed separation will only increase as long as the escaping particles are sufficiently close to the peak of the distribution and have a sufficiently high density and then decrease again as the tail stretches out and its density decreases below the observable value. The time between the collision and the peak of the separation is typically comparable to the dynamical time of the sub-system. We conclude that, in contrast to the conventional expectation, the separation is \emph{harder} to observe in more evolved systems.

Using a simple analytical model of rare self-interactions, we find that in general smaller and more tightly bound systems will exhibit less separation than larger and more loosely bound systems. Moreover, in more tightly bound systems, the typical time-scales (in particular the dynamical time) are shorter, meaning that it will take a shorter time to ``undo the damage'' caused by DM self-interactions.

We therefore conclude that despite of their smaller surface densities, A520 and the Musket Ball Cluster should be as suitable as the Bullet Cluster in constraining or measuring rare DM self-interactions using the separation between DM and galaxies and they may even be more suitable than the Bullet Cluster for probing effective drag forces. However, for a scattering cross-section of $\sigma = 1$--$2\ \text{cm}^2\,\text{g}^{-1}$, we find typical separations (between the centroids of the areas containing 68\% of the total DM and galaxy mass) to be 20--40~kpc, which is \emph{below} the current level of sensitivity. Nevertheless, as astronomers discover more and more such mergers and map their gravitational potentials using weak lensing, such small separations may become measurable.

\section*{Acknowledgements}

We thank Celine Boehm, Will Dawson, John Magorrian, Paul McMillan, Paolo Panci, Andrew Pontzen, Prasenjit Saha and Liliya Williams for discussions and valuable comments on the manuscript. MTF acknowledges a `Sapere Aude' Grant no.\ 11-120829 from the Danish Council for Independent Research. FK is supported by the Studienstiftung des Deutschen Volkes, STFC UK, and a Leathersellers' Company Scholarship at St Catherine's College, Oxford. SS acknowledges support by the EU Marie Curie Initial Training Network `UNILHC' (PITN-GA-2009-237920) and a Niels Bohr Professorship award from the Danish National Research Foundation. KSH acknowledges partial support by the German Science Foundation (DFG) under the Collaborative Research Center (SFB) 676 `Particles, Strings and the Early Universe'. MTF and FK thank the KITP for hospitality during the workshop \emph{Hunting for Dark Matter}, where part of this project was carried out. This research was supported in part by the National Science Foundation under Grant No. NSF PHY11-25915.

\appendix

\section{Evaporation and deceleration rates for frequent self-interactions}
\label{ap:evaporation}

In this appendix we calculate the evaporation and deceleration rate arising from frequent DM self-interactions for a DM halo $S_1$ moving through a larger system $S_2$. For the purpose of this appendix it will be convenient to assume that $S_2$ is at rest and the DM particles in $S_1$ move at a uniform velocity $\bmath{v} = \bmath{v}_0$. For a single collision, we define $\delta v = |\bmath{v}' - \bmath{v}|$ and $\delta v_\parallel = (\bmath{v}' - \bmath{v}) \cdot \bmath{v}_0 / v_0$. In terms of the scattering angle $\theta$ we find 
\begin{align}
\delta v & = v_0 \sin \theta = v_0 \sin \left(\theta_\text{cms}/2 \right)\\ 
\delta v_\parallel & = v_0 \sin^2 \theta = v_0 \sin^2 \left( \theta_\text{cms}/2 \right) \; .
\end{align}
As expected, we find for small scattering angles $\delta v \gg \delta v_\parallel$, implying that the change of velocity is largest in the direction perpendicular to the initial velocity $\bmath{v}_0$. However, over the course of many collisions these perpendicular changes will average to zero. What remains non-zero over a large number of collisions is the average of $\delta v^2$ and the average of $\delta v_\parallel$. The first term corresponds to an increase of energy of the DM particles, leading to cumulative evaporation, the second term corresponds to a deceleration of the halo.

In a time $\mathrm{d}t$ we have 
\begin{equation}
\mathrm{d}C = \frac{\rho_2}{m_\text{DM}} \, \frac{\mathrm{d}\sigma}{\mathrm{d}\Omega} \, v_0 \, \mathrm{d}t \, \mathrm{d}\Omega
\end{equation}
collisions and therefore:
\begin{align}
\mathrm{d}(v^2) & = 2 \pi \, \frac{\rho_2 \, v_0^3 \, \mathrm{d}t}{m_\text{DM}} \int \frac{\mathrm{d}\sigma}{\mathrm{d}\Omega_\text{cms}} \, \sin^2 \frac{\theta_\text{cms}}{2} \, \sin \theta_\text{cms} \, \mathrm{d}\theta_\text{cms} \\ 
\mathrm{d}v_\parallel & = 2 \pi \, \frac{\rho_2 \, v_0^2 \, \mathrm{d}t}{m_\text{DM}} \int \frac{\mathrm{d}\sigma}{\mathrm{d}\Omega_\text{cms}} \, \sin^2 \frac{\theta_\text{cms}}{2} \, \sin \theta_\text{cms} \, \mathrm{d}\theta_\text{cms} \; .
\end{align}
Note that, by assumption, the contribution of expulsive collisions is negligible, so we do not have to exclude them explicitly from the range of integration. Also, we observe that the two expressions are completely identical apart from a factor $v_0$. Defining the rate of cumulative evaporation as $R_\text{cml} \equiv \dot{E} / E \approx v_0^{-2} \, \mathrm{d} (v^2) / \mathrm{d}t$ and the deceleration rate as $R_\text{dec} \equiv v_0^{-1} \, \mathrm{d} v_\parallel / \mathrm{d}t$, we therefore find
\begin{equation}
R_\text{cml} \approx R_\text{dec} \approx \pi \, \frac{\rho_2 \, v_0}{m_\text{DM}} \int_{-1}^1 \frac{\mathrm{d}\sigma}{\mathrm{d}\Omega_\text{cms}} \, (1 - \cos \theta_\text{cms}) \, \mathrm{d} \cos \theta_\text{cms} \; .
\end{equation}

In the formula above we have neglected the fact that for $\theta_\text{cms} > \pi/2$ the roles of the two DM particles will be exchanged and the particle originally considered part of $S_1$ will belong to $S_2$ after the collision (and vice versa). To take this effect into account, we restrict the range of integration to $\cos \theta_\text{cms} \in \left[0, 1\right]$ and include an additional factor of 2 to write
\begin{align}
R_\text{cml} \approx R_\text{dec} & \approx 2 \pi \, \frac{\rho_2 \, v_0}{m_\text{DM}} \int_0^1 \frac{\mathrm{d}\sigma}{\mathrm{d}\Omega_\text{cms}} \, (1 - \cos \theta_\text{cms}) \, \mathrm{d} \cos \theta_\text{cms} \nonumber \\ 
& = \frac{\rho_2 \, v_0 \, \sigma_\mathrm{T}}{2 \, m_\text{DM}} \; .
\end{align}
In the last step, we have defined the \emph{momentum transfer cross-section} for indistinguishable particles
\begin{equation}
\sigma_\text{T} \equiv 4 \pi \int_0^{1} \mathrm{d}\cos\theta_\text{cms} \, (1 - \cos \theta_\text{cms}) \,
\frac{\mathrm{d}\sigma}{\mathrm{d}\Omega_\text{cms}} \; ,
\end{equation}
which determines the rate at which energy and momentum is transferred
from $S_2$ to $S_1$.\footnote{Note that the conventional definition of $\sigma_\text{T}$ overestimates the momentum transfer for $\theta_\text{cms} > \pi/2$ because it does not take into account the fact that the two DM particles are indistinguishable and can therefore always be relabelled in such a way that one particle scatters with $\theta_\text{cms} < \pi/2$. Our definition of $\sigma_\text{T}$ differs in that we integrate $\theta_\text{cms}$ only from $0$ to $\pi/2$ and include an additional factor of 2. For isotropic scattering, our definition gives $\sigma_\text{T} = \sigma/2$, while the conventional definition gives $\sigma_\text{T} = \sigma$.}

Let us briefly review the assumptions we have made in the derivation of our results. Most importantly, we have assumed that all particles in $S_2$ are at rest and all particles in $S_1$ move with the common
velocity $v_0$. More realistically, these particles can have a range of velocities described by the density functions $f_1(\bmath{v}_1,\bmath{x}_1)$ and $f_2(\bmath{v}_2,\bmath{x}_2)$, respectively, where $\bmath{v}_i$ and $\bmath{x}_i$ are measured relative to the centre of mass of $S_i$. The distribution of relative velocities $v_\text{rel}$ will be given by $f_\text{rel}(\bmath{v}_\text{rel}) = \int \mathrm{d}^3 v_1
f_1(\bmath{v}_1) \, f_2(\bmath{v}_\text{rel} - \bmath{v}_1 - \bmath{v}_0)$. The rate at which a particle from $S_1$ encounters particles from $S_2$ is then given by
\begin{equation}
\mathrm{d}C = \frac{\rho_2}{m_\text{DM}} \, \frac{\mathrm{d}\sigma}{\mathrm{d}\Omega} \, v \, f_\text{rel}(\bmath{v}) \, \mathrm{d}^3 v \, \mathrm{d}t \, \mathrm{d}\Omega \; .
\end{equation}
For a more accurate treatment, we should therefore replace $v_0$ by $f_\text{rel}(\bmath{v})$ in our results for $R_\text{cml}$ and $R_\text{dec}$ and integrate over $\mathrm{d}^3 v$. However, as long as $v_0$ is large compared to typical values for $v_1$ and $v_2$, our approximate results above are sufficient.

\section{Examples for frequent self-interactions}
\label{ap:frequent}

In this appendix we give details about two examples for differential cross-sections that can lead to frequent DM self-interactions, i.e.~interactions with a very small fraction of expulsive collisions. For $2\rightarrow2$ scattering, the scattering amplitude can depend on the incoming and outgoing momenta only via the Lorentz-invariant Mandelstam variables $s$, $t$ and $u$. For non-relativistic scattering these are $s = 4 \, m_\text{DM}^2 \left(1 + v_\text{cms}^2\right)$, $t = -2 \, m_\text{DM}^2 \, v_\text{cms}^2 \left(1 - \cos \theta_\text{cms}\right)$ and \mbox{$u = -2 \, m_\text{DM}^2 \, v_\text{cms}^2 \left(1 + \cos \theta_\text{cms}\right)$}. Moreover, since we assume that the two DM particles are indistinguishable, the cross-section must be symmetric in $t$ and $u$. The simplest possible combinations are $t + u$, $t^2 + u^2$ and $t \, u$. Out of these three terms, only the third exhibits a strong angular dependence. Since we are interested in self-interactions with predominantly small momentum transfer corresponding to small scattering angles, we will focus on cross-sections containing inverse powers of $t \, u$, which diverge in the limit $\theta_\text{cms} \rightarrow 0$ and $\theta_\text{cms} \rightarrow \pi$.\footnote{Note that a scattering angle $\theta_\text{cms} \approx \pi$ does not lead to large momentum transfer but simply exchanges the roles of the two particles.}

\subsection{Long-range interactions}

First of all, we consider the case where DM particles interact with each other via the exchange of a massless vector mediator (such as a dark photon~\citep{Ackerman:2008gi, Feng:2009mn}). For non-relativistic scattering we can approximate the differential cross-section by
\begin{align}
\frac{\mathrm{d}\sigma}{\mathrm{d}\Omega} & = \frac{\alpha'^2}{s} \left[\frac{s^2+u^2}{t^2}+\frac{s^2+t^2}{u^2}+\frac{2s^2}{t u}\right] \nonumber \\
& \approx \alpha'^2 \frac{s (t+u)^2}{(t\,u)^2}  
\nonumber
\\
&= \frac{\alpha'^2}{m_\text{DM}^2 \, v_\text{cms}^4 \, (1-\cos^2\theta_\text{cms})^2} \; , \label{eq:CSlong}
\end{align}
where $\alpha'$ is the coupling strength of the vector mediator. Typical parameters of interest are $\alpha' = 10^{-2}$ for $m_\text{DM} = 1\,\text{TeV}$~\citep{Ackerman:2008gi}. By construction, the cross-section in Eq.~(\ref{eq:CSlong}) diverges for $\theta \rightarrow 0$ and $\theta \rightarrow \pi$, so we need to introduce a cut-off at small and large angles. We take~\citep{CyrRacine:2012fz}
\begin{equation}
\theta_\text{min} = \frac{4 \, \alpha'}{\lambda_\text{De} \, m_\text{DM} \, v_0^2} \; , \quad \theta_\text{max} = \pi - \theta_\text{min}
\end{equation}
where the Debye screening length is given by
\begin{equation}
\lambda_\text{De} = \frac{1}{4} \frac{m_\text{DM} \, v_0}{\sqrt{\pi \, \alpha' \, \rho}}
\end{equation}
with $\rho$ the total density of DM particles. Even for very large DM densities, e.g.~close to the centre of a DM halo, we find $\lambda_\text{De} \gg 1\,\text{m}$, implying that $\theta_\text{min}$ is a very small number.

First of all, we can calculate the rate of expulsive collisions. For $\kappa = 2 \, v_\text{esc}^2 / v_0^2$ we find
\begin{equation}
f \approx \theta_\text{min}^2 \left[\frac{1 - \kappa}{(2 - \kappa) \, \kappa} + \text{atanh}(1 - \kappa)\right] \; ,
\end{equation}
which is tiny because of the smallness of $\theta_\text{min}$. The corresponding rate of immediate evaporation is
\begin{equation}
R_\text{imd} = \frac{32 \pi \, \alpha'^2 \, \rho_2}{m_\text{DM}^3 \, v_0^3} \left[\frac{1 - \kappa}{(2 - \kappa) \, \kappa} + \text{atanh}(1 - \kappa)\right] \; ,
\end{equation}
where we have used $v_\text{cms} \approx v_0/2$. For $0.2 \la \kappa \la 0.8$ the expression in square brackets gives values in the range $0.5$--$3$. For the cumulative evaporation, on the other hand, we find
\begin{equation}
R_\text{cml} = \frac{8 \pi \, \alpha'^2 \, \rho_2}{m_\text{DM}^3 \, v_0^3} \left[1 - 2 \log \left(\theta_\text{min}/2\right) \right] \; .
\end{equation}
For typical parameter choices, the square brackets give values of $\mathcal{O}(10^2)$. This logarithmic enhancement of the cumulative evaporation rate implies $R_\text{cml} \gg R_\text{imd}$. As desired, we can therefore have a large deceleration rate without being constrained by immediate evaporation. We can write the decelerating force as
\begin{align}
\frac{F_\text{drag}}{m_\text{DM}} = v_0 \, R_\text{dec} & = \frac{8 \pi \, \alpha'^2 \, \rho_2}{m_\text{DM}^3 \, v_0^2} \left[1 - 2 \log \left(\theta_\text{min}/2\right) \right] \nonumber \\ 
& = \frac{\tilde{\sigma}}{4 \, m_\text{DM}} \frac{\rho_2}{v_0^2} \; ,
\end{align}
where we have defined the effective cross-section
\begin{equation}
\frac{\tilde{\sigma}}{m_\text{DM}} = \frac{32 \pi \, \alpha'^2}{m_\text{DM}^3} \left[1 - 2 \log \left(\theta_\text{min}/2\right) \right] \; .
\end{equation}
This result is in complete analogy to the case of dynamical friction, where $\alpha'$ is replaced by $G_\text{N} \, M$ and $\log \theta_\text{min}/2$ is replaced by the Coulomb logarithm $\log \Lambda$. Because of the factor $v_0^{-2}$, the effects of this drag force will be most interesting for slowly moving systems.

Before we move on to the next example, let us discuss the case where the mediator has a small but non-zero mass. In this case $\lambda_\text{De}$ in the definition of $\theta_\text{min}$ is replaced by the de Broglie wavelength of the dark photon $\lambda = 1/m_{A'}$. Nevertheless, as long as the mediator is weakly coupled, i.e.~$m_{A'} \ll m_\text{DM} v^2 / \alpha'$, we still have $f \ll 1 $ and $R_\text{imd} < R_\text{cml}$.

\subsection{Velocity-independent interactions}

We have seen above that long-range interactions satisfy the requirement that immediate evaporation can be neglected. However, the velocity dependence of the cross-section implies that these interactions are generally of little interest for galaxy clusters. In this section, we explore the possibilities to have frequent interactions without a suppression of large velocities. If we want to have a differential cross-section that does not depend on the relative velocity $v_0$, we need to have equal powers of $t$ and $u$ in the numerator and denominator. As an example, we consider the case\footnote{We leave the discussion of possible particle physics models leading to such a cross-section to future work. As a convenient parameterisation we assume that~-- as in the case of long-range interactions~-- the cross-section is proportional to $\alpha'^2 / m_\text{DM}^2$.}
\begin{equation}
\frac{\mathrm{d}\sigma}{\mathrm{d}\Omega_\text{cms}} = \frac{\alpha'^2}{s} \frac{t^2 + u^2}{t \, u} \approx \frac{\alpha'^2 \,}{2 \, m_\text{DM}^2} \frac{1+\cos^2\theta_\text{cms}}{1-\cos^2\theta_\text{cms}} \; .
\end{equation}
While the total cross-section diverges, the momentum transfer cross-section is finite:
\begin{equation}
\sigma_\text{T} = \frac{\alpha'^2}{m_\text{DM}^2} \pi \left(\log(16)-1\right) \; .
\end{equation}

We find
\begin{equation}
f = \frac{1 - \kappa - 2 \, \text{atanh}(1 - \kappa)}{1 + 2 \log (\theta_\text{min} / 2)} \; .
\end{equation}
For sufficiently small $\theta_\text{min}$, we find as expected $f \ll 1$. However, since there is one less power of $t \, u$ in the
denominator than in the case of long-range interactions, we now find
\begin{align}
R_\text{imd} & = 2 \pi \, \rho_2 \, v_0 \frac{\alpha'^2}{m_\text{DM}^3} \left[ \kappa - 1 + 2 \, \text{atanh}(1 - \kappa) \right] \\
R_\text{cml} & = \frac{\pi}{2} \rho_2 \, v_0 \frac{\alpha'^2}{m_\text{DM}^3} (\log 16 - 1)
\end{align}
implying that $R_\text{imd} \ga R_\text{cml}$ for $\kappa \la 0.6$. In other words, in spite of the small number of expulsive collisions, the immediate evaporation rate is comparable to the cumulative evaporation rate. Consequently, we have to take both effects into account when comparing to observational bounds. 

To conclude this section, we note that the resulting drag force is given by
\begin{equation}
\frac{F_\text{drag}}{m_\text{DM}} = \frac{\alpha'^2}{m_\text{DM}^3} \frac{\pi}{2} \rho_2 \, v_0^2 (\log 16 - 1) = \frac{\tilde{\sigma}}{4\,m_\text{DM}} \rho_2 \, v_0^2 \; .
\end{equation}
with\footnote{We have chosen the normalisation of $\tilde{\sigma}$ in such a way that $\sigma_\mathrm{T} = \tilde{\sigma}/2$ in analogy to the case of isotropic scattering.}
\begin{equation}
\frac{\tilde{\sigma}}{m_\text{DM}} = \frac{\alpha'^2}{m_\text{DM}^3} 2 \pi (\log 16 - 1) \; .
\end{equation}
Since the fundamental interactions are independent of velocity, the effective drag force is proportional to $v_0^2$. Such a force is expected to arise from any velocity-independent self-interaction cross-section, provided the fraction of expulsive collisions is sufficiently small so that we can average over a large number of interactions. While we considered a particularly simple form for the differential cross-section here, we expect to obtain a similar effective drag force also for more complicated cases. 

\subsection{Observational constraints}

\subsubsection*{Long-range interactions}

Because of the strong velocity dependence of long-range interactions, we expect the strongest constraints to arise from systems with low velocities. For example, we can obtain a bound by requiring Carina, Draco and Ursa Minor to survive until the present day (see~\citet{Gnedin:2000ea}). In other words, we require that the typical timescale of evaporation caused by the motion of these objects through the Milky Way DM halo is sufficiently larger than the age of the Milky Way, i.e.
\begin{equation}
R_\text{cml}^{-1} > 10^{10}\,\text{yr} \; .
\end{equation}
The background DM density is of the order of $10^{-26}\,\text{g}\,\text{cm}^{-3}$ and the relative velocity of the dwarf spheroidals is roughly $150\,\text{km\,s}^{-1}$~\citep{Gnedin:2000ea}. This translates to
\begin{equation}
 \frac{\alpha'^2}{m_\text{DM}^3} \la 10^{-11}\,\text{GeV}^{-3} \; , \quad \frac{\tilde{\sigma}}{m_\text{DM}} \la 10^{-11}\,\text{cm}^2\,\text{g}^{-1} \; .
\end{equation}
The corresponding drag force is then constrained to be
\begin{equation}
\frac{F_\text{drag}}{m_\text{DM}} \la 10^{-12}\,\text{m s}^{-2} \left(\frac{v_0}{100\,\text{km\,s}^{-1}}\right)^{-2} \left(\frac{\rho_2}{0.01\,\text{GeV\,cm}^{-3}}\right) \; .
\end{equation}
We note that even stronger constraints on $\alpha'^2 / m_\text{DM}^3$ have been obtained by studying galaxy ellipticity~\citep{Feng:2009mn}. However, these bounds apply only under the assumption that the hot gas is in hydrostatic equilibrium and has negligible rotation (see~\citet{Binney1990,Buote2002, Diehl2007, Brighenti2009}). We thus prefer the more conservative bounds from evaporation.

\subsubsection*{Velocity-independent interactions}

For velocity-independent interactions the drag force grows with velocity, so we expect strong constraints from the Bullet Cluster, which has $v_0 \approx 4500\,\text{km\,s}^{-1}$. Following~\citet{Markevitch:2003at}, we derive a constraint by requiring that the sub-cluster loses no more than $\Delta N / N < 30\%$ of its mass during the collision. We integrate $R_\text{imd}$ and $R_\text{cml}$ along the trajectory of the sub-cluster to find
\begin{equation}
\frac{\Delta N}{N} \approx \int (R_\text{imd} + R_\text{cml}) \, \mathrm{d}t \approx 1 - \exp\left[-8 \, \Sigma_2 \frac{\alpha'^2}{m_\text{DM}^3} \right] \; .
\end{equation}
Estimating $\Sigma = 0.3 \, \text{g\,cm}^{-2}$ for the Bullet Cluster~{\citep{Markevitch:2003at}}, we obtain the constraint
\begin{equation}
\frac{\alpha'^2}{m_\text{DM}^3} \la 550\,\text{GeV}^{-3} \; , \quad \frac{\tilde{\sigma}}{m_\text{DM}} \la 1.2\,\text{cm}^2\,\text{g}^{-1} \; .
\end{equation}
corresponding to
\begin{equation}
\frac{F_\text{drag}}{m_\text{DM}} < 10^{-9}\,\text{m\,s}^{-2} \left(\frac{v_0}{4500\,\text{km\,s}^{-1}}\right)^2 \left(\frac{\rho_2}{1\,\text{GeV\,cm}^{-3}}\right) \; .
\end{equation}

\section{Details on the numerical simulations}
\label{ap:numerical}

\subsection{Systems under consideration}

We perform numerical simulations for three different systems which are chosen to be representative of known major mergers. System A consists of two galaxy clusters of similar mass ($M = 1.5 \times 10^{14} \,
\mathrm{M}_{\sun}$ and $M = 3.5 \times 10^{14} \, \mathrm{M}_{\sun}$) with a relative velocity of $v_0 \sim 2400\,\text{km\,s}^{-1}$. We describe these clusters with Hernquist profiles~\citep{Hernquist}\footnote{We prefer the Hernquist profile over an NFW profile for the cluster haloes as it has a finite mass, a finite central potential and an analytical expression for the velocity distribution function. Since our results are largely independent of the behaviour of the haloes at large radii, both profiles give nearly identical results.} 
\begin{equation}
\rho(r) = \frac{M \, b}{2 \pi \, r (r + b)^3}
\end{equation}
with $b = 200\,\text{kpc}$ and $b = 400\,\text{kpc}$, respectively. These parameters are chosen in such a way as to resemble the colliding clusters in A520~\citep{Mahdavi:2007yp, Girardi:2008jp, Jee:2012sr, Clowe:2012am}.

System B consists of two galaxy clusters of very different mass, similar to the Bullet Cluster~\citep{Barrena:2002dp,Clowe:2003tk,Markevitch:2003at,Clowe:2006eq}. For the larger cluster we take a Hernquist profile with $b = 1000\,\text{kpc}$ and $M =  2.5 \times 10^{15} \, \mathrm{M}_{\sun}$, for the smaller cluster we take $M = 3 \times 10^{14} \, \mathrm{M}_{\sun}$ and $b = 100\,\text{kpc}$~\citep{Randall:2007ph}. We assume a collision velocity of $4500\,\text{km\,s}^{-1}$, even though it has been argued that the velocity of the Bullet Cluster may be significantly smaller~\citep{Springel:2007tu}.

Finally, our System C is representative of a merger with small velocity such as the Musket Ball Cluster~\citep{2012ApJ...747L..42D, Dawson:2012fx}. We model both haloes by Hernquist profiles with slightly different mass, using $M = 3 \times 10^{14} \, \mathrm{M}_{\sun}$ and $b = 400\,\text{kpc}$ for the first halo and $M = 1.5 \times 10^{14} \, \mathrm{M}_{\sun}$ and $b = 300\,\text{kpc}$ for the second halo. The collision velocity is $2000\,\text{km\,s}^{-1}$. For simplicity, we will always refer to the larger cluster as the main cluster and the smaller cluster as the sub-cluster, even when both clusters are similar in size. Note that we do not attempt a precise matching of our simulations to observed cluster collisions as in~\citet{Randall:2007ph}. Rather, our Systems A--C are meant to illustrate how our conclusions can vary for different systems. 

In our simulations, the parameters of the gravitational potential of the main cluster are taken to be time-independent, only the position and velocity of the main cluster can change in the gravitational field of the sub-cluster. For the sub-cluster, on the other hand, central density and scale radius are time-dependent and are determined self-consistently from the simulation. To initialise the simulation, the program randomly chooses a representative set of DM particles and galaxies bound to the sub-cluster, using the known density profile and velocity distribution for the initial parameters specified above. At each time step, the program then calculates the motion of all these particles in the combined gravitational field of the main cluster and the sub-cluster. At the end of each step, the gravitational potential of the sub-cluster is updated by evaluating the current position of all DM particles in the simulation. For this purpose, it is assumed that the sub-cluster can be described by a Hernquist profile at all times. This is consistent with the observational constraint that DM self-interactions cannot change the profile of a DM halo over timescales as short as $10^8\,\text{yr}$. However, our approach neglects the gravitational pull of particles which leave the DM halo. Since this additional force will affect DM and galaxies in the same way, it does not significantly influence the separation between DM halo and galaxies (see \S~\ref{sec:separation}).

We assume that the sub-cluster starts from rest and falls towards the main cluster, so the collision occurs with negligible impact parameter. For non-zero impact parameter, the sub-cluster would probe a smaller integrated DM density and therefore be less affected by DM self-interactions. We have not studied such cases in detail and refer to~\citet{Randall:2007ph} for a discussion.

\subsection{Implementation of self-interactions}

\subsubsection*{Frequent self-interactions}
In our simplified approach we can easily model an effective drag force on the DM halo by including an additional contribution to the acceleration of DM particles. This contribution is proportional to the DM self-interaction cross-section, the DM density of the main cluster and to $v_\text{rel}^2$, where $v_\text{rel}$ is the velocity of the DM particle under consideration relative to the main cluster.

\subsubsection*{Rare self-interactions}

In order to simulate rare self-interactions rather than an effective drag force, we need to calculate the probability for each DM particle to scatter based on its velocity, the surrounding density of DM particles and the scattering cross-section. A random number generator is then used to decide whether scattering does occur and how it will affect the velocity of the DM particle. This is done separately for the DM density of the main cluster and for the DM density of the sub-cluster as we will now discuss.

The probability for a DM particle at position $\bmath{x}$ and with velocity $\bmath{v}$ relative to the main cluster to scatter on the main cluster within a time $\mathrm{d}t$ is $\mathrm{d}p = \rho_2(x) \, v_\text{rel} \, \mathrm{d}t \, \sigma / m_\text{DM}$, where $v_\text{rel} = |\bmath{v} - \bmath{v}_2|$ and $\bmath{v}_2$ is randomly chosen from the velocity distribution of the main cluster, which we assume to be a Maxwell-Boltzmann distribution.\footnote{Using the velocity distribution corresponding to a Hernquist profile gives very similar results while being computationally far more expensive.} We take $v_\text{dis} = 1000\,\text{km\,s}^{-1}$, $1200\,\text{km\,s}^{-1}$ and $800\,\text{km\,s}^{-1}$ for Systems A, B and C, respectively. If a collision occurs, the velocity of the particle changes in such a way that the scattering is isotropic in the centre-of-mass frame.

If the momentum transfer in the collision is large, i.e.~comparable to the initial momenta of the two DM particles, we now need to trace the motion of both particles, i.e.~we need to generate a new particle corresponding to the DM particle that was previously part of the main system (and therefore not included in our simulation), but has now received a large momentum transfer. This is necessary, because it is possible that the particle from the main cluster loses so much energy in the collision that it actually becomes bound to the sub-cluster. Neglecting this contribution would therefore overestimate the evaporation rate. Note that this implies that the number of particles in the simulation increases with time. This is accounted for by reducing the density of the main cluster accordingly. On the other hand, once a DM particle has left the sub-cluster, further collisions will no longer be of interest for us because they do not affect our observables. We therefore neglect the scattering of DM particles which are far away from the sub-cluster (meaning that their distance is large compared to both the scale radius and the tidal radius of the sub-cluster).

The probability for a DM particle to scatter with another particle from the sub-cluster is given by $\mathrm{d}p = \rho_1(x) \, v_\text{rel} \, \mathrm{d}t \, \sigma / m_\text{DM}$, where $v_\text{rel} = |\bmath{v} - \bmath{v}_1|$ and $\bmath{v}_1$ is randomly chosen from the velocity distribution of the sub-cluster (determined by the parameters of the Hernquist profile). If a scattering process does occur, the program determines the simulated DM particle closest to the original particle, determines their CM frame and randomises the direction of the velocities in that frame.

Most of the collisions between DM particles in the sub-cluster will occur in the central region of the halo with small momentum transfer. While these collisions are interesting for observables like halo shapes, they are not relevant for the separation between the DM halo and galaxies. Moreover, for the values of the self-interaction cross-section and the timescales that we are interested in, no significant changes in the halo profile are expected. Therefore, we will only consider the scattering between two DM particles within the sub-cluster if at least one of them has already scattered with large momentum transfer in the past.

The remaining concern is that in the presence of large DM self-interactions the central densities are reduced and the cuspy halo profiles that we use will no longer be a good approximation. Indeed, the observed separation depends on the central density of the main cluster (see~\citet{Randall:2007ph}). Since we do not aspire to make precise predictions for the separation, we neglect the impact of the density profile of the main cluster and assume that it can approximately be described by a Hernquist profile even in the presence of self-interactions.

\section{An analytical description of rare self-interactions}
\label{ap:analytical}

In this appendix we derive a simple model to describe the separation between a DM halo and galaxies resulting from rare self-interactions. We shall see that~-- in spite of its simplicity~-- the model is sufficient to reproduce the main features observed in our numerical simulations (see Fig.~\ref{fig:results} and Fig.~\ref{fig:analytical}). The idea is to trace the orbits of scattered particles assuming that all collisions take place at the centre of the DM halo and that the bound particles are initially at rest. After colliding with a DM particle with velocity $\bmath{v}_0$, the probability distribution of velocities of an originally bound DM particle is $f_\text{cms}(\bmath{v}_\text{cms}) = \delta ( v_\text{cms}^2 - v_\text{0}^2 / 4) / (\pi v_\text{0})$ in the CM frame, corresponding to $f(\bmath{v}) = \delta ( v^2 - \bmath{v} \cdot \bmath{v}_0) / (\pi v_\text{0})$ in the sub-cluster frame.

A DM particle travelling with a velocity $v$ immediately after the collision will subsequently slow down in the gravitational potential $\Phi(\bmath{x})$ of the sub-cluster.\footnote{We neglect the gravitational potential of the main cluster in our model.} Its distance from the centre will then be given by $r(t, v)$, which is defined implicitly via
\begin{equation}
t = \int_0^{r(t,v)} \frac{\mathrm{d}r}{\sqrt{v^2 + 2 (\Phi(0) - \Phi(r))}} \; .
\end{equation}
Using the initial distribution of velocities, we can then calculate the distribution of positions of the DM particle after a time $t$:
\begin{equation}
f(\bmath{x}, t) = \int_{v < v_\text{crit}(t)} \frac{\delta (r - r(t,v) ) \, \delta(\theta_r - \theta_v)}{2\pi \, \sin \theta_r \, r^2} \, f(\bmath{v}) \, \mathrm{d}^3v \; ,
\end{equation}
where $v_\text{crit}(t)$ is the cut-off velocity, meaning that all particles with initial velocity $v > v_\text{crit}(t)$ are considered to have left the halo at time $t$. The decreasing number of particles in the halo is reflected by the fact that $\int f(\bmath{x}, t) \, \mathrm{d}^3 x = \text{min}(v_\text{crit}(t)^2 / v_0^2, \, 1)$.

Ultimately, we are interested in the displacement in the $z$-direction, i.e.~in the direction parallel to $v_0$. To calculate the expectation value we can write
\begin{equation}
\langle z(t) \rangle = \int z f(\bmath{x}, t) \mathrm{d}^3 x = \frac{I(t)}{v_0^3}
\label{eq:zexp}
\end{equation}
with
\begin{equation}
I(t) = \int_{0}^{v_\text{crit}(t)} 2\, r(t, v) \, v^2 \, \mathrm{d}v \; .
\end{equation}
We can think of $\langle z(t) \rangle$ as the response function of the system. It gives the separation caused by a collision at a time $t$ in the past. To calculate the total separation resulting from the collision of two clusters, we need to integrate this expression along the path of the system.

If the distance between the two clusters is given by $x(t)$ and their relative velocity by $v_0(t)$, the flux of DM particles as a function of time is given by $j(t) = v_0(t) \, \rho_2(x(t)) / m_\text{DM}$, and the infinitesimal probability of an interaction in time $\mathrm{d}t$ is $\mathrm{d}p(t) = \sigma \, j(t) \, \mathrm{d}t$. The total separation at time $t$ is therefore given by
\begin{align}
\Delta z(t) & = \frac{\sigma}{m_\text{DM}} \int \langle z(t') \rangle \, v_0(t - t') \, \rho_2(x(t - t')) \, \mathrm{d}t' \nonumber  \\
& = \frac{\sigma}{m_\text{DM}} \int \frac{I(t')}{v_0(t - t')^2} \, \rho_2(x(t - t')) \, \mathrm{d}t' \; .
\end{align}
In principle, this equation can be evaluated numerically for arbitrary DM density and velocity profiles. Here we choose to employ a very simple profile, namely the isochrone model, for which we can calculate all orbits analytically. The isochrone model is defined by the potential~\citep{Binney2008}
\begin{equation}
\Phi(r) = -\frac{G_\text{N} M}{b + \sqrt{b^2 + r^2}} \; .
\end{equation}
We choose the parameters of the isochrone model in such a way as to match the simulated Hernquist profiles at large radii.\footnote{At small radii, the density of the isochrone model is significantly lower, leading to a central potential that is shallower than for the Hernquist model. This partially compensates for neglecting the initial kinetic energy of the DM particles before scattering. Moreover, the isochrone model has the advantage that all quantities remain finite for $r\rightarrow0$, $v\rightarrow0$.} For the main cluster, we employ the same density profile as in the simulations (see Appendix~\ref{ap:numerical} and Table~\ref{tab:parameters}) and take $\rho_2$ to be the average density within a distance $b$ from the path of the sub-cluster. 

As in our simulations, we do not want to include particles too far away from the sub-cluster. We can achieve this goal in an approximate way by specifying the cut-off velocity $v_\text{crit}(t)$. First of all, we only want to include particles with $v \cos \theta < v_0 / 2$, or equivalently $v < v_0 / \sqrt{2}$. Particles with higher velocity should be considered part of the main cluster rather than part of the sub-cluster. This requirement removes exactly half of the particles. However, we have neglected the fact that two particles participate in each collision. If one of the particles is initially at rest, there will always be exactly one particle with $v < v_0 / \sqrt{2}$ after the collision. We should therefore multiply our expression for $\Delta z(t)$ by a factor of 2 to reflect the fact that for large momentum transfer the incoming DM particle can be caught by the sub-cluster.

Furthermore, we do not want to include particles far away from the sub-cluster. We therefore require $z(v, t) = r(t, v) v/v_0 < z_\text{max}$ and take $z_\text{max} \equiv 3b/2$.\footnote{While $z_\text{max}$ should be comparable to the size of the DM halo, there is some arbitrariness in the precise definition. In practice, we find better agreement with the numerical simulations for $z_\text{max} = 3b/2$ than for $z_\text{max} = b$.} Since the left-hand side depends only on $v$ and $t$, this inequality can be rewritten in the form $v < v_\text{max}(t)$, where $v_\text{max}(t)$ is the initial velocity of a particle that needs time $t$ to travel a distance $z_\text{max}$ in $z$-direction. Note that $v_\text{max}$ decreases with increasing $t$, because as the system evolves particles with smaller initial velocity will reach the critical distance $z(v,t) = z_\text{max}$. We then define $v_\text{crit}(t) = \text{min}(v_0 / \sqrt{2}, v_\text{max}(t))$. Using this definition of $v_\text{crit}(t)$ we calculate the separation predicted for System B. Our results are shown in Fig.~\ref{fig:analytical} and are in good agreement with the separation obtained by our numerical simulation.

\bigskip

To conclude this section, let us make some rough estimates of the typical time and distance scales over which the separation develops. At small times after the cluster collision, the separation is dominated by particles with $v \sim v_\text{esc}$ (see Fig.~\ref{fig:analytical}). For a Hernquist profile, the averaged escape velocity is given by $v_\text{esc} \sim \sqrt{G_\text{N} \, M / (3 \, b)}$ where $M$ is the mass and $b$ the typical size of the sub-cluster. Neglecting the deceleration in the gravitational potential of the sub-cluster, particles with $v \sim v_\text{esc}$ will take approximately the time
\begin{equation}
t_\text{esc} \sim \frac{z_\text{max}}{v_\text{esc}} \sim (2\text{--}2.5)\,\sqrt{\frac{b^3}{G_\mathrm{N} \, M}}  
\end{equation}
to leave the sub-cluster. The separation should therefore peak approximately at a distance $z = v_0 \, t_\text{esc}$, which gives $z \sim 600,\text{kpc}$, $z \sim 300\,\text{kpc}$ and $z \sim 1000\,\text{kpc}$ for Systems A, B and C~-- in agreement with our numerical simulations.

A description of the behaviour of the system at later times is more involved and we restrict ourselves to making an estimate of the time for scattered particles that remain bound to the sub-cluster to return to its central region. The typical time-scale for an orbit is given by the dynamical time $t_\text{dyn} = 1/\sqrt{G_\text{N}\,\bar{\rho}_1}$, where $\bar{\rho}_1$ is the average density within the orbit. As a simple estimate, we take a Hernquist profile and calculate the average density for $r < b$, which yields $\bar{\rho}_1 \approx 0.25 \, M / b^3$, giving $t_\text{dyn} \sim t_\text{esc}$. The typical time to complete half an orbit is then
\begin{equation}
t_\text{orb} \approx \pi \, t_\text{dyn} \sim (6\text{--}8) \,\sqrt{\frac{b^3}{G_\mathrm{N} \, M}} \; .
\end{equation}
For $t > t_\text{orb}$ we expect particles that have remained bound in the cluster collision to give a negative contribution to the total separation. Note that for such large times the deceleration of the sub-cluster in the gravitational field of the main cluster can no longer be neglected, so that $z_\text{neg} < v_0 \, t_\text{neg}$.

To estimate the maximum magnitude of the separation, we simply assume that at $t \sim t_\text{esc}$ all escaping particles are roughly at a distance $b$ from the main cluster and we neglect the contribution of non-expulsive collisions. Using the fraction of expulsive collisions, we then find
\begin{equation}
\Delta z \sim \frac{f \, \Sigma_2 \, \sigma \, b}{m_\text{DM}} \; .
\end{equation}
For a cross-section $\sigma / m_\text{DM} = 1.6\,\text{cm}^2\,\text{g}^{-1}$ this estimate gives a separation of $\Delta z \sim 20 \,\text{kpc}$, $\Delta z \sim 30 \,\text{kpc}$ and $\Delta z \sim 15 \,\text{kpc}$ for Systems A, B and C~-- in good agreement with our numerical simulations. The larger values of $b$ in Systems A and C compensate for the smaller surface density $\Sigma_2$, so the expected maximum separation in all three systems is comparable.

\end{document}